\documentclass[11pt]{article} % For LaTeX2e
\usepackage{fullpage}

\usepackage{comment}
\usepackage{url}
\usepackage{amsmath}
\usepackage{graphicx,xspace}

\usepackage{paralist}
\usepackage{algorithm,algorithmic}
\newfloat{Algorithm}{tb}{lox}
\floatname{Algorithm}{Algorithm}

\usepackage[colorlinks,linkcolor=red,citecolor=blue,urlcolor=blue,draft]{hyperref}
\usepackage{caption}
\usepackage{subcaption}
\DeclareCaptionType{copyrightbox}
\usepackage{tabularx}
\usepackage{etoolbox}
\usepackage{mdwlist}
\usepackage{breakurl}

\usepackage{tikz}
\usepackage{pgfplots}
\usetikzlibrary{shapes,shapes.gates.logic.US,trees,positioning,arrows,automata,decorations.pathmorphing,chains}
\usetikzlibrary{shadows,calc,backgrounds,patterns,fit}
\usetikzlibrary{matrix,arrows}
\colorlet{darkgreen}{green!50!black}
\colorlet{darkred}{red!90!black}
\colorlet{darkyellow}{yellow!90!black}

\tikzset{
  gnode/.style={
    fill=white,
    draw=black,
    circle,
    very thick, 
    inner sep=3.5,
    drop shadow={shadow xshift=0.3ex,shadow yshift=-0.5ex, path
      fading={circle with fuzzy edge 20 percent}}
  }
}

\tikzset{
  rnode/.style={
    fill=black,
    draw=black,
    circle,
    very thick, 
    inner sep=3.5,
    drop shadow={shadow xshift=0.3ex,shadow yshift=-0.5ex, path
      fading={circle with fuzzy edge 20 percent}}
  }
}

\tikzset{
  ynode/.style={
    fill=black!50!white,
    draw=black,
    circle,
    very thick, 
    inner sep=3.5,
    drop shadow={shadow xshift=0.3ex,shadow yshift=-0.5ex, path
      fading={circle with fuzzy edge 20 percent}}
  }
}

\input{Definitions}

\def\yahoolda{{\sf Yahoo!\ LDA}\xspace}
\def\yahooldaD{{\sf Yahoo!\ LDA(D)}\xspace}
\def\yahooldaM{{\sf Yahoo!\ LDA(M)}\xspace}

\title{A Scalable Asynchronous Distributed Algorithm for Topic Modeling}
\author{
Hsiang-Fu Yu\\
       {University of Texas, Austin}\\
       {rofuyu@cs.utexas.edu}
\and
Cho-Jui Hsieh\\
        {University of Texas, Austin}\\
        {cjhsieh@cs.utexas.edu}
\and
Hyokun Yun\\
        {Amazon}\\
        {yunhyoku@amazon.com}
\and
S.V.N Vishwanathan\\
        {University of California, Santa Cruz}\\
        {vishy@ucsc.edu}
\and
Inderjit S. Dhillon\\
        {University of Texas, Austin}\\
        {inderjit@cs.utexas.edu}
}
\def\gridwidth{2}
\def\nodenum{4}
\def\asnwidth{0.5}
\def\asnmargin{0.05}

\newcommand{\asnbox}[3]{
  \fill [#3!25!lightgray]
  (#1 * \asnwidth + \asnmargin, 
  \gridwidth * 3 - #2 * \gridwidth + \asnmargin)
  rectangle
  (#1 * \asnwidth + \asnwidth - \asnmargin, 
  \gridwidth * 3 - #2 * \gridwidth + \gridwidth - \asnmargin);
  \draw
  (#1 * \asnwidth + \asnmargin, 
  \gridwidth * 3 - #2 * \gridwidth + \asnmargin)
  rectangle
  (#1 * \asnwidth + \asnwidth - \asnmargin, 
  \gridwidth * 3 - #2 * \gridwidth + \gridwidth - \asnmargin);
}

\newcommand{\dasnbox}[2]{
  \draw[dashed]
  (#1 * \asnwidth + \asnmargin, 
  \gridwidth * 3 - #2 * \gridwidth + \asnmargin)
  rectangle
  (#1 * \asnwidth + \asnwidth - \asnmargin, 
  \gridwidth * 3 - #2 * \gridwidth + \gridwidth - \asnmargin);
}

\newcommand{\dtpt}[2]{
  \draw (#1 * \asnwidth + 0.5 * \asnwidth, 
  #2 * \asnwidth + 0.5 * \asnwidth) 
  node {\tiny{x}};
}

\newcommand{\mvarr}[4]{
  \draw [->]
  (#1 * \asnwidth + 0.5 * \asnwidth, 
  3 * \gridwidth - #2 * \gridwidth + 0.5 * \gridwidth) 
  --
  (#3 * \asnwidth + 0.5 * \asnwidth, 
  3 * \gridwidth - #4 * \gridwidth + 0.5 * \gridwidth);
}

\newcommand{\boxlayout}{
  \draw (0,0) rectangle (\gridwidth * \nodenum, \gridwidth * \nodenum);

  \fill [red!10] (0, 3 * \gridwidth) 
  rectangle (\gridwidth * \nodenum, 4 * \gridwidth);

  \fill [green!10] (0, 2 * \gridwidth) 
  rectangle (\gridwidth * \nodenum, 3 * \gridwidth);

  \fill [blue!10] (0, 1 * \gridwidth) 
  rectangle (\gridwidth * \nodenum, 2 * \gridwidth);

  \fill [brown!10] (0,0) 
  rectangle (\gridwidth * \nodenum, 1 * \gridwidth);

  \foreach \y in {1,2,3}
  {
    \draw[densely dashed] 
    (-\gridwidth * 0.1, \gridwidth * \y)
    --
    (\gridwidth * \nodenum + \gridwidth * 0.1,  \gridwidth * \y);
  }

  \dtpt{5}{1}  \dtpt{0}{15}  \dtpt{13}{12}  \dtpt{13}{6}  \dtpt{14}{8}  \dtpt{0}{8}  \dtpt{7}{10}  \dtpt{15}{2}  \dtpt{10}{5}  \dtpt{8}{13}  \dtpt{5}{3}  \dtpt{12}{5}  \dtpt{10}{7}  \dtpt{11}{6}  \dtpt{15}{3}  \dtpt{6}{9}  \dtpt{13}{5}  \dtpt{1}{3}  \dtpt{8}{15}  \dtpt{1}{6}  \dtpt{1}{7}  \dtpt{8}{9}  \dtpt{12}{4}  \dtpt{12}{1}  \dtpt{13}{10}  \dtpt{9}{0}  \dtpt{8}{4}  \dtpt{15}{12}  \dtpt{3}{6}  \dtpt{10}{13} 
}

\newcommand{\ptrna}{crosshatch}
\newcommand{\ptrnb}{north west lines}
\newcommand{\ptrnc}{north east lines}
\newcommand{\ptrnd}{horizontal lines}

\newcommand{\ptcla}{red}
\newcommand{\ptclb}{green}
\newcommand{\ptclc}{blue}
\newcommand{\ptcld}{brown}

\newcommand{\drawrowpart}{
  \draw[pattern=\ptrnd, pattern color=\ptcld] 
  (-\gridwidth - \gridwidth * 0.25, 0 * \gridwidth) 
  rectangle 
  (- \gridwidth * 0.25, 
  0 * \gridwidth + 1 * \gridwidth);
  \draw[pattern=\ptrnc, pattern color=\ptclc] 
  (-\gridwidth - \gridwidth * 0.25, 
  1 * \gridwidth) 
  rectangle 
  (- \gridwidth * 0.25, 
  1 * \gridwidth + 1 * \gridwidth);
  
  \draw[pattern=\ptrnb, pattern color=\ptclb] 
  (-\gridwidth - \gridwidth * 0.25, 
  2 * \gridwidth) 
  rectangle 
  (- \gridwidth * 0.25, 
  2 * \gridwidth + 1 * \gridwidth);
  
  \draw[pattern=\ptrna, pattern color=\ptcla] 
  (-\gridwidth - \gridwidth * 0.25, 
  3 * \gridwidth) 
  rectangle 
  (- \gridwidth * 0.25, 
  3 * \gridwidth + 1 * \gridwidth);
}

\newcommand{\drawcolpart}[4]{
  \draw[pattern=#3, pattern color=#4] 
  (#1 * \asnwidth, 
  4 * \gridwidth + 0.25 * \gridwidth) 
  rectangle 
  (#2 * \asnwidth + \asnwidth, 
  4 * \gridwidth + 0.25 * \gridwidth + \gridwidth);
}

\newcommand{\Ft}{{\mathtt F}}
\begin{document}

% make the title area
\maketitle

\begin{abstract}
  Learning meaningful topic models with massive document collections
  which contain millions of documents and billions of tokens is
  challenging because of two reasons: First, one needs to deal with a
  large number of topics (typically in the order of thousands). Second,
  one needs a scalable and efficient way of distributing the computation
  across multiple machines. In this paper we present a novel algorithm
  F+Nomad LDA which simultaneously tackles both these problems. In order
  to handle large number of topics we use an appropriately modified
  Fenwick tree. This data structure allows us to sample from a
  multinomial distribution over $T$ items in $O(\log T)$ time. Moreover,
  when topic counts change the data structure can be updated in
  $O(\log T)$ time. In order to distribute the computation across
  multiple processor we present a novel asynchronous framework inspired
  by the Nomad algorithm of \cite{YunYuHsietal13}. We show that F+Nomad LDA
  significantly outperform state-of-the-art on massive problems which
  involve millions of documents, billions of words, and thousands of
  topics. 
\end{abstract}

%\category{I.2.6}{Artificial Intelligence}{Leraning}
%\terms{Algorithms, Experimentation}
%\keywords{Topic Models; Scalability; Sampling} % NOT required for Proceedings

\section{Introduction}

Topic models provide a way to aggregate vocabulary from a document
corpus to form latent ``topics.'' In particular, Latent Dirichlet
Allocation (LDA) \cite{BleNgJor03} is one of the most popular topic
modeling approaches.  Learning meaningful topic models with massive
document collections which contain millions of documents and billions of
tokens is challenging because of two reasons: First, one needs to deal
with a large number of topics (typically in the order of
thousands). Second, one needs a scalable and efficient way of
distributing the computation across multiple machines.

% When applying LDA to large document corpora, the task of inferring the
% latent topics is challenging as the number of variables is very large.

% However, thanks to the use of a conjugate prior
% in the model, one can integrate out the natural parameters and derive 
% an efficient collapsed Gibbs sampler (CGS) \cite{GriSte04}. 

% In a very
% similar vein, one can also derive an efficient collapsed variational
% Bayes (CVB0) sampler \cite{AsuWelSmyTeh09}. Although CGS and CVB0 are
% very efficient inference approaches for LDA, the inherent sequential
% nature of the updates makes them unattractive for massive text
% datasets. 

% With our increasing ability to store and process text data, it is
% increasingly commonplace to find document collections of a few hundreds
% of Gigabytes. Therefore, 

Unsurprisingly, there has been significant resources devoted to
developing scalable inference algorithms for LDA. To tackle large number
of topics, \cite{YaoMimMcC09} proposed an ingenious sparse sampling
trick that is widely used in packages like MALLET and Yahoo! LDA. More
recently, in an award winning paper, \cite{LiAhmRavSmo14} proposed using
the alias table method to speed up sampling from the multinomial
distribution. On the other hand, there has also been significant effort
towards distributing the computation across multiple processors. Some
early efforts in this direction include the work of \cite{YanXuQi09} and
\cite{IhlNew12}. The basic idea here is to partition the documents
across processors. During each inner iteration the words in the
vocabulary are partitioned across processors and each processor only
updates the latent variables associated with the subset of documents and
words that it owns. After each inner iteration, a synchronization step
is used to update global counts and to re-partition the words across
processors. In fact, a very similar idea was discovered in the context
of matrix completion independently by \cite{GemNijHaaSis11} and
\cite{RecRe13}. However, in the case of LDA we need to keep a global
count synchronized across processors which significantly complicates
matters as compared to matrix completion. Arguably, most of the recent
efforts towards scalable LDA such as \cite{SmoNar10,NewAsuSmyWel09} have
been focused on this issue either implicitly or explicitly. Recently
there is also a growing trend in machine learning towards asynchronous
algorithms which avoid bulk synchronization after every iteration. In
the context of LDA see the work of \cite{AsuSmyWel08}, and in the more
general machine learning context see \eg,
\cite{GonLowGuBicetal12,LiAndParSmoetal14}.

In this paper, we propose F+Nomad LDA which simultaneously tackles the
twin problems of large number of documents and large number of topics.
In order to handle large number of topics we use an appropriately
modified Fenwick tree. This data structure allows us to sample from a
multinomial distribution over $T$ items in $O(\log T)$ time. Moreover,
when topic counts change the data structure can be updated in
$O(\log T)$ time. In order to distribute the computation across multiple
processor we present a novel asynchronous framework inspired by the
Nomad algorithm of \cite{YunYuHsietal13}. While we believe that our framework can
handle variable update schedules of many different methods, in this
paper we will primarily focus on Collapsed Gibbs Sampling (CGS). Our
technical contributions can be summarized as follows:
% Inspired by \cite{YunYuHsietal13}, we propose Nomad LDA an asynchronous
% distributed framework for topic models, in particular LDA. Our framework
% can handle the variable update schedules of many different inference
% methods. In this paper, however, we will focus on CGS and CVB0. Our
% technical contributions can be summarized as follows:
% Nomad LDA, for a generic approach to parallelize
% various inference methods for LDA.  We analyze a common variable access
% structure shared by many inference methods such as collapsed Gibbs
% sampling and collapsed variational Bayes and identify a few properties
% of this structure, which give us a better understanding about the
% potential parallelism behind this access pattern.
\begin{itemize}
\item We identify the following key property of various inference
  methods for topic modeling: only a single vector of size $k$ needs to
  be synchronized across multiple processors.
\item We present a variant of of the Fenwick tree which allows us to
  efficiently encode a multinomial distribution using $O(T)$
  space. Sampling can be performed in $O(\log T)$ time and maintaining
  the data structure only requires $O(\log T)$ work. 
\item F+Nomad LDA: A novel parallel framework for various types of
  inference methods for topic modeling. Our framework utilizes the
  concept of nomadic tokens to avoid locking and conflict at the same
  time. Our parallel approach is fully asynchronous with non-blocking
  communication, which leads to good speedups. Moreover, our approach
  minimizes the staleness of the variables (at most $k$ variables can be
  stale) for distributed parallel computation.
\item We demonstrate the scalability of our methods by performing
  extensive empirical evaluation on large datasets which contain
  millions of documents and \textbf{billions} of words.
\end{itemize}

\section{Notation and Background}
\label{sec:NotationBackground}

We begin by very briefly reviewing Latent Dirichlet Allocation (LDA)
\cite{BleNgJor03}. Suppose we are given $I$ documents denoted as
$d_{1}, d_{2}, \ldots, d_{I}$, and let $J$ denote the number of words in
the vocabulary.  Moreover, let $n_{i}$ denote the number of words in a
document $d_{i}$. Let $w_j$ denote the $j$-th word in the vocabulary and 
$w_{i,j}$ denote the $j$-th word in the $i$-th document. Assume that the 
documents are generated by sampling from $T$ 
topics denoted as $\phi_{1}, \phi_{2},\ldots, \phi_{T}$; a topic is
simply a $J$ dimensional multinomial distribution over words. Each
document includes some proportion of the topics.  These proportions are
latent, and we use the $T$ dimensional probability vector $\theta_{i}$
to denote the topic distribution for a document $d_{i}$. Moreover, let
$z_{i,j}$ denote the latent topic from which $w_{i,j}$ was drawn. Let
$\alpha$, and $\beta$ be hyper parameters of the Dirichlet
distribution. The generative process for LDA can be described as
follows:
\begin{enumerate}
\item Draw $T$ topics $\phi_{k} \sim \mathtt{Dirichlet}(\beta)$.
\item For each document $d_{i} \in \{d_{1}, d_{2}, \ldots, d_{I}\}$:
  \begin{itemize}
  \item Draw $\theta_{i} \sim \mathtt{Dirichlet}(\alpha)$.
  \item For each word $w_{i,j}$ with $j = 1, \ldots, n_{i}$
    \begin{itemize}
    \item Draw $z_{i,j} \sim \mathtt{Discrete}(\theta_{i})$.
    \item Draw $w_{i,j} \sim \mathtt{Discrete}(\phi_{z_{i,j}})$.
    \end{itemize}
  \end{itemize}
\end{enumerate}

\subsection{Inference}
\label{sec:Inference}
Collapsed Gibbs Sampling (CGS) \cite{GriSte04} is a popular inference scheme
for LDA. %Let $\gamma_{z,i,j,w} := I\rbr{z_{i,j} = z \text{ and } w_{i,j} = w}$,  
%Two popular inference schemes for LDA namely collapsed Gibbs sampling
%(CGS) \cite{GriSte04} and collapsed variational Bayes (CVB0)
%\cite{AsuWelSmyTeh09} can be viewed in a unified framework. Towards
%this end, 
Define  
\begin{equation}
  \label{eq:ndef}
  n_{z, i, w} := \sum_{j=1}^{n_{i}} I\rbr{z_{i,j} = z \text{ and } w_{i,j} = w},  
\end{equation}
%and 
%where $\gamma_{z,i,j,w} := I\rbr{z_{i,j} = z \text{ and } w_{i,j} = w}$,  
$n_{z,i,*}
= \sum_{w} n_{z,i,w}$, $n_{z,*,w} = \sum_{i} n_{z, i, w}$, and $n_{z, *,
  *} = \sum_{i, w} n_{z, i, w}$. The update rule for CGS can be
written as follows
\begin{enumerate}
\item Decrease $n_{z_{i,j}, i, *}$, $n_{z_{i,j}, *, w_{i,j}}$, and
  $n_{z_{i,j}, *, *}$ by $1$.
\item Resample $z_{i,j}$ according to 
  \begin{align}
\Pr\rbr{z_{i,j} | w_{i,j}, \alpha,
      \beta}
      \propto \frac{ \rbr{n_{z_{i,j}, i, *} + \alpha_{z_{i,j}}}
      \rbr{n_{z_{i,j}, *, w_{i,j}}+ \beta_{w_{i,j}}}}{n_{z_{i,j}, *, *}+
      \sum_{j=1}^{J} \beta_{j}}.
    \label{eq:cgs-update}
  \end{align}
\item Increase $n_{z_{i,j}, i, *}$, $n_{z_{i,j}, *, w_{i,j}}$, and
  $n_{z_{i,j}, *, *}$ by $1$.
\end{enumerate}
Although in this paper we will focus on CGS, note that there are many
other inference techniques for LDA such as collapsed variational Bayes,
Stochastic Variational Bayes, or Expectation Maximization which
essentially follow a very similar update
pattern~\cite{AsuWelSmyTeh09}. We believe that the parallel framework
proposed in this paper will apply to this wider class of inference
techniques as well.

\subsection{Review of Multinomial Sampling}
Given a $T$-dimension discrete distribution characterized by 
unnormalized parameters $\pb$ with $p_t\ge0$ such as in 
\eqref{eq:cgs-update}, many sampling algorithms can be  
applied to draw a sample $z$ such that $\Pr(z=t) \propto p_{t}$. 
\begin{itemize}
  \item {LSearch:} Linear search on $\pb$. {\bf Initialization.} Compute the 
    normalization constant $c_T = \sum_{t} p_t$. {\bf Generation.} First 
    generate $u=\mathtt{uniform}(c_T)$, a uniformly random number in $[0,c_T)$, and 
    perform a linear search to find $z = \min \cbr{t: \rbr{\sum_{s\le t} p_s} > u}$. 
  \item {BSearch:} Binary search on $\cbb = \mathtt{cumsum}(\pb)$ {\bf 
    Initialization.} Compute $\cbb = \mathtt{cumsum}(\pb)$ such that 
    $c_t=\sum_{s:s\le t} p_s$.  
    {\bf Generation.} First generate $u=\mathtt{uniform}(c_T)$ and perform a 
      binary search on $\cbb$ to find $z = \min \cbr{t: c_t > u}$. 
    \item {Alias method.} {\bf Initialization.} Construct an Alias table
      \cite{Walker77} for $\pb$, which contains two arrays of length
      $T$: $alias$ and $prob$.  See \cite{vose91} for an linear time
      construction scheme. {\bf Generation.} First generate
      $u=\mathtt{uniform}(T)$, $j=\floor{u}$, and
    \begin{equation*}
   z = \begin{cases}
      j+1 &  \text{if } (u-j) \le prob[j+1] \\
      alias[j+1] & \text{o.w.}
    \end{cases}.
    \end{equation*}
\end{itemize}
See Table \ref{tab:disc_samplers} for a comparison of the time/space
requirements of each of the above sampling methods. 
\begin{table*}
  \caption{Comparison of samplers for a $T$-dimensional multinomial 
    distribution $\pb$ described by unnormalized parameters 
    $\cbr{p_t: t = 1,\dots,T}$. }
  \label{tab:disc_samplers}
  \begin{center}
  \resizebox{1\linewidth}{!}{
    \begin{tabular}{ll||cc||c||c}
      & \multicolumn{1}{c}{Data Structure} & \multicolumn{2}{c}{Initialization} & \multicolumn{1}{c}{Generation} & Parameter Update\\ 
      & \multicolumn{1}{c||}{Space} & Time & Space & Time & Time \\
      \hline
      LSearch & $c_T = \pb^T\one$: $\Theta(1)$ & $\Theta(T)$ & $\Theta(1)$ &  $\Theta(T)$ & $\Theta(1)$\\
      BSearch &$\cbb= \mathtt{cumsum}(\pb)$: $\Theta(T)$ & $\Theta(T)$ & $ \Theta(1)$ & \multicolumn{1}{c||}{$\Theta(\log T)$} & $\Theta(T)$\\
      Alias Method & $prob, alias$: $\Theta(T)$ & $\Theta(T)$ & $\Theta(T)$ & \multicolumn{1}{c||}{$\Theta(1)$} & $\Theta(T)$\\ 
      F+tree Sampling & $\Ft.\mathtt{initialize}(\pb)$: $\Theta(T)$& $\Theta(T)$ & $\Theta(1)$ & \multicolumn{1}{c||}{$\Theta(\log T)$} & $\Theta(\log T)$
    \end{tabular}
    }
  \end{center}
\end{table*}

\tikzset{
  normal/.style = {label = center:},
  head/.style = {fill = orange!90!blue,
                 label = center:},
  tail/.style = {fill = blue!70!yellow, text = black,
                 label = center:\textsf{\Large T}}
}

\begin{figure*}
  \begin{center}
    \hspace{-2em}
    \begin{subfigure}[t]{0.4\textwidth}
      \centering      
      \begin{tikzpicture}[
          scale = 0.6, transform shape, thick,
          every node/.style = {draw, circle, minimum size = 10mm},
          normal/.style = {label = center:},
          head/.style = {fill = orange!90!blue, label = center:},
          tail/.style = {fill = blue!70!yellow, text = black, label = center:\textsf{\Large T}}
          grow = down,  % alignment of characters
          level 1/.style = {sibling distance=5cm},
          level 2/.style = {sibling distance=2.5cm}, 
          %level 3/.style = {sibling distance=2cm}, 
          level distance = 2cm,
          label distance = -1mm
        ]
        \node[label={[name=Startlabel] \sffamily\large 001},draw, normal] (Start) { 2.5 }
        child { node [label={[name=Alabel] \sffamily\large 010}, normal] (A) {1.8}
          child { node [label={\sffamily\large 100},draw, normal] (B) {0.3}}
          child { node [label={\sffamily\large 101}, normal] (C) {1.5}}
        }
        child { node [label={[name=Dlabel] \sffamily\large 011}, normal] (D) {0.7}
          child { node [label={\sffamily\large 110},draw, normal] (E) {0.4}}
          child { node [label={\sffamily\large 111}, normal] (F) {0.3}}
        };
        % Labels
        \begin{scope}[nodes = {draw = none}]
          \node[right=11pt] at (F) {\color{blue} $\bf =p_4$}; 
          \node[right=11pt] at (C) {\color{blue} $\bf =p_2$}; 
          \node[right=11pt] at (E) {\color{blue} $\bf =p_3$}; 
          \node[right=11pt] at (B) {\color{blue} $\bf =p_1$}; 

          \node[right=11pt] at (A) {\color{blue} \bf =0.3+1.5}; 
          \node[right=11pt] at (D) {\color{blue} \bf =0.4+0.3}; 
          \node[right=11pt] at (Start) {\color{blue} \bf =1.8+0.7}; 

          \begin{scope}[nodes = {below = 11pt}]
            \node at (B) {\large $1$};
            \node at (C) {\large $2$};
            \node at (E) {\large $3$};
            \node at (F) {\large$4$};
          \end{scope}
          \node [draw, densely dashed, rectangle, rounded corners, 
          thin, fit=(A)(D)(Start)(Alabel)(Dlabel)(Startlabel)]{};
        \end{scope}
      \end{tikzpicture}
      \caption{F+tree for $\pb=[0.3, 1.5, 0.4, 0.3]^T$}
      \label{fig:btree-plot-costruct}
    \end{subfigure}
    \smallskip
    \begin{subfigure}[t]{0.3\textwidth}
      \centering
      \begin{tikzpicture}[
          scale = 0.6, transform shape, thick,
          every node/.style = {draw, circle, minimum size = 10mm},
          normal/.style = {label = center:},
          head/.style = {fill = orange!90!blue, label = center:},
          tail/.style = {fill = blue!70!yellow, text = black, label = center:\textsf{\Large T}}
          grow = down,  % alignment of characters
          level 1/.style = {sibling distance=5cm},
          level 2/.style = {sibling distance=2.5cm}, 
          %level 3/.style = {sibling distance=2cm}, 
          level distance = 2cm,
          label distance = -1mm
        ]
        \node[label={[name=Startlabel] \sffamily\large 001}, head, draw, very thick] (Start) { 2.5 }
        child { node [label={[name=Alabel] \sffamily\large 010},  normal] (A) {1.8}
          child { node [label={\sffamily\large 100},  normal] (B) {0.3}}
          child { node [label={\sffamily\large 101},  normal] (C) {1.5}}
        }
        child { node [label={[name=Dlabel] \sffamily\large 011}, head, draw, very thick] (D) {0.7} edge from parent[draw,very thick,->]
          child { node [label={\sffamily\large 110},  head] (E) {0.4}}
          child { node [label={\sffamily\large 111},  normal,draw, thick] (F) {0.3} edge from parent[draw, thick, -]}
        };
        % Labels
        \begin{scope}[nodes = {draw = none}]
          \node[left=6pt] at (Start.west) {\color{blue}\large \bf u=2.1};
          \node[right=6pt] at (Start.east) {\color{red}\large \bf u$\ge$1.8 $\searrow$};
          \node[left=13pt] at (D) {\color{blue}\large \bf u=0.3}; 
          \node[right=1pt] at (D.east) {\color{red}\large \bf u$<0.4$ $\swarrow$};

          \begin{scope}[nodes = {below = 11pt}]
            \node at (B) {\large $1$};
            \node at (C) {\large $2$};
            \node at (E) {\large $3$};
            \node at (F) {\large $4$};
          \end{scope}
            \node [draw, densely dashed, rectangle, rounded corners, 
            thin, fit=(A)(D)(Start)(Alabel)(Dlabel)(Startlabel)]{};
        \end{scope}
      \end{tikzpicture}
      \caption{Sampling}
      \label{fig:btree-plot-sample}
    \end{subfigure}
    \quad
    \begin{subfigure}[t]{0.3\linewidth}
      \centering
      \begin{tikzpicture}[
          scale = 0.6, transform shape, thick,
          every node/.style = {draw, circle, minimum size = 10mm},
          normal/.style = {label = center:},
          head/.style = {fill = orange!90!blue, label = center:},
          tail/.style = {fill = blue!70!yellow, text = black, label = center:\textsf{\Large T}}
          grow = down,  % alignment of characters
          level 1/.style = {sibling distance=5cm},
          level 2/.style = {sibling distance=2.5cm}, 
          %level 3/.style = {sibling distance=2cm}, 
          level distance = 2cm,
          label distance = -1mm
        ]
        \node[label={[name=Startlabel] \sffamily\large 001}, head] (Start) { 3.5 }
        child { node [label={[name=Alabel] \sffamily\large 010},  normal] (A) {1.8}
           child { node [label={\sffamily\large 100},  normal] (B) {0.3}} 
           child { node [label={\sffamily\large 101},  normal] (C) {1.5}}
         }
         child { node [ label={[name=Dlabel] \sffamily\large 011}, head, draw, very thick] (D) {1.7}
           child { node [label={\sffamily\large 110},  head] (E) {1.4} edge from parent[draw,very thick,<-] }
           child { node [label={\sffamily\large 111},  normal, draw, thick] (F) {0.3} 
             edge from parent[draw, thick,-]} 
             edge from parent[draw,very thick,<-]
         };
        % Labels
        \begin{scope}[nodes = {draw = none}]
          \node[right=11pt] at (E) {\bf =0.4\color{red}+$\delta$};
          \node[right=11pt] at (D) { \bf =0.7\color{red}+$\delta$};
          \node[right=11pt] at (Start) {\bf =2.5\color{red}+$\delta$}; 

          \begin{scope}[nodes = {below = 11pt}]
            \node at (B) {\large $1$};
            \node at (C) {\large $2$};
            \node at (E) {\large $3$};
            \node at (F) {\large$4$};
          \end{scope}
          \node [draw, densely dashed, rectangle, rounded corners, 
          thin, fit=(A)(D)(Start)(Alabel)(Dlabel)(Startlabel)]{};
        \end{scope}
      \end{tikzpicture}
      \caption{Updating (with $\delta=1.0$)}
      \label{fig:btree-plot-update}
    \end{subfigure}
  \end{center}
  \caption{Illustration of sampling and updating using F+tree in logarithmic time.}
  \label{fig:btree-plot}
\end{figure*}
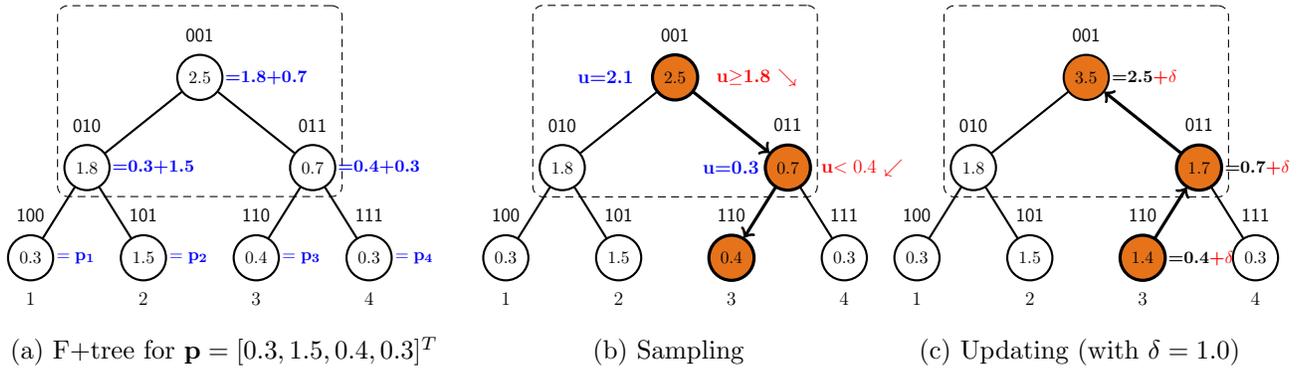

\section{Fenwick Tree Sampling}
\label{sec:ProposedSampling}
In this section, we first describe a binary tree structure F+tree for fast 
$T$-dimensional multinomial sampling.  
The initialization of an F+tree is linear in $T$ and the cost to generate a sample
is logarithmic to $T$. Furthermore, F+tree can also be maintained in  
logarithmic time for a single parameter update on $p_t$. Due to the 
efficiency on both sample generation and parameter updates, we will 
explain how F+tree sampling can be used to accelerate LDA sampling 
significantly.  

\subsection{F+tree Sampling}
\label{sec:treesampling}
F+tree, first introduced for weighted sampling without replacement~\cite{WongEaston80}, is a 
simplified and generalized version of Fenwick tree~\cite{Fenwick1994}, which 
supports both efficient sampling and update procedures. In fact, Fenwick tree 
can be regarded as a compression version of F+tree studied in this paper. For 
the simplicity, we assume $T$ is a power of  
2. F+tree is a complete binary tree with $2T-1$ nodes for a given $\pb$, where
\begin{itemize}
  \item each leaf node corresponds to a dimension $t$ and stores $p_t$ as its value, and
  \item each internal node stores the sum of the values of all of its leaf descendant, or 
    equivalently the sum of values of its two children due to binary tree 
    structure.  
\end{itemize}
See Figure \ref{fig:btree-plot-costruct} for an example with
$\pb=[0.3, 1.5, 0.4, 0.3]$ and $T=4$.  Nodes in the dotted rectangle are
internal nodes.  Similar to the representation used in
Heap~\cite{CorLeiRiv90}, an array $\Ft$ of length $2T$ can be used to
represent the F+tree structure.  Let $i$ be the index of each node, and
$\Ft[i]$ be the value stored in the $i$-th node. The index of the left
child, right child, and parent of the $i$-th node is $2i$, $2i+1$, and
$i/2$, respectively. The 0/1 string along each node in
\ref{fig:btree-plot} is the binary number representation of the node
index.
\par
{\bf Initialization.}
By the definition of F+tree, given $\pb$, the values of $\Ft$ be defined as follows.
\begin{equation}
  \Ft[i] = 
  \begin{cases}
    p_{i-T+1} & \text{if } i \ge T, \\
    \Ft[2i] + \Ft[2i+1] & \text{if } i < T.
  \end{cases}
  \label{eq:btree-def}
\end{equation}
Thus, $\Ft$ can be constructed in $\Theta(T)$ by reversely initializing 
elements using \eqref{eq:btree-def}. Unlike Alias method, in addition to $\Ft$, there is 
no extra space required in the F+tree initialization. 
\par
{\bf Sample Generation.} Sampling on a F+tree can be carried out as a 
simple top-down traversal procedure to locate 
$z = \min\cbr{t: \rbr{\sum_{s:s\le t} p_s} > u}$ for  
a number uniformly sampled between $[0,\sum_{t}p_t)$. Note that 
$\sum_{t} p_t$ is stored in $\Ft[1]$, which can be directly used to generate 
$u=\mathtt{uniform}(\Ft[1])$. Let $\mathtt{leaves}(i)$ be the set of all leaf 
descendant of the $i$-th node. We can consider a general recursive step in the 
traversal with the current node $i$ and $u\in[0,\Ft[i])$. The 
definition of F+tree guarantees that 
\begin{align*}
  u \ge \Ft[i.\mathtt{left}] &\Rightarrow z\in \mathtt{leaves}(i.\mathtt{right}),\\
  u < \Ft[i.\mathtt{left}] &\Rightarrow z \in \mathtt{leaves}(i.\mathtt{left}),
\end{align*}
This provides a guideline to determine which child to go next. If right child 
is chosen, $\Ft[i.\mathtt{left}]$ should be subtracted from $u$ to ensure 
$u \in [0, \Ft[i.\mathtt{right}])$. 
Note that as half of $t$s are removed from the set of candidate, it is clear 
that this sampling procedure costs only $\Theta(\log T)$ time. The detailed 
procedure, denoted by $\Ft.\mathtt{sample}(u)$, is described in Algorithm 
\ref{alg:tree-sampling}. A toy example  
with initial $u=2.1$ is illustrated in Figure \ref{fig:btree-plot-sample}. 
\begin{algorithm}[htbp]
  \begin{compactitem}
    \item[] Input: $\Ft$: an F+tree for $\pb$, $u = \mathtt{uniform}(\Ft[1])$.
    \item[] Output: $z = \min \cbr{t: \rbr{\sum_{s\le t} p_s} > u}$
    \item $i\leftarrow 1$
    \item While $i$ is not a leaf
      \begin{compactitem}
      \item If $u \ge \Ft[i.\mathtt{left}]$, 
        \begin{compactitem}
        \item $u \leftarrow u - \Ft[i.\mathtt{left}]$
        \item $i \leftarrow i.\mathtt{right}$
        \end{compactitem}
      \item Else
        \begin{compactitem}
        \item $i \leftarrow i.\mathtt{left}$
        \end{compactitem}
      \end{compactitem}
    \item $z \leftarrow i-T+1$
  \end{compactitem}
  \caption{Logarithmic time sampling: $\Ft.\mathtt{sample}(u)$.}
  \label{alg:tree-sampling}
\end{algorithm}
\begin{algorithm}[htbp]
  \begin{compactitem}
    \item[] Input: a F+tree $\Ft$ for $\pb$, $t$, $\delta$.
    \item[] Output: F+tree $\Ft$ is updated for $\bar{\pb}\equiv\pb + \delta\eb_t$
    \item $i \leftarrow \mathtt{leaf}[t]$
    \item While $i$ is a valid node
      \begin{compactitem}
        \item $\Ft[i] = \Ft[i]+\delta$
        \item $i \leftarrow i.\mathtt{parent}$
      \end{compactitem}
  \end{compactitem}
  \caption{Logarithmic time F+tree maintenance for a single parameter update: 
  $\Ft.\mathtt{update}(t, \delta)$}
  \label{alg:tree-update}
\end{algorithm}

\par
{\bf Maintenance for Parameter Updates.} 
A simple and efficient maintenance routine to deal with slight changes on the 
multinomial parameters $\pb$ can be very useful in CGS for LDA (See details in 
Section \ref{sec:treelda}). F+tree structure supports a logarithmic time 
maintenance routine for a single element change on $\pb$.
Assume the $t$-th component is updated by $\delta$:  
\[
\bar{\pb} \leftarrow \pb +\delta \eb_t,
\]
where $\eb_t$ is the $t$-th column of the identity matrix of order $T$. 
A simple bottom-up update procedure to modify a F+tree $\Ft$ for the current 
$\pb$ to a F+tree for $\bar{\pb}$ can be carried out as follows. Let 
$\mathtt{leaf}[t]$ be the leaf node corresponding to $t$. For all the 
ancestors $i$ of $\mathtt{leaf}[t]$ (self included), perform the following 
delta update: 
\[
\Ft[i] = \Ft[i] + \delta.  
\]
See Figure \ref{fig:btree-plot-update} for an illustration with $t=3$ and 
$\delta=1.0$.  The detailed procedure, denoted by 
$\Ft.\mathtt{update}(t,\delta)$, is described in Algorithm 
\ref{alg:tree-update}.  The maintenance cost is linear to the depth of the 
F+tree, which is $\Theta(\log T)$. Note that to deal with the similar change on $\pb$, 
LSearch can update its normalization constant $c_T \leftarrow c_T + \delta$ in 
a constant time, while both BSearch and Alias method require to re-construct
the entire data structure (either $\cbb=\mathtt{cumsum}(\pb)$ or the Alias  
table: $alias$ and $prob$), which costs $\Theta(T)$ time in general.   

See Table \ref{tab:disc_samplers} for a summary of complexity analysis for 
each multinomial sampling approach. Clearly, LSearch has the smallest
update cost but the largest generation cost, and Alias method has the 
best generation cost but the worst maintenance cost. In contrast, F+tree 
sampling has a logarithmic time procedure for both operations.

\begin{algorithm}[htbp]
  \begin{compactitem}
    \item $\Ft.\mathtt{initialize}(\qb)$, with $q_t = \frac{\beta}{n_t + \bar{\beta}}$
    \item For each word $w$ 
      \begin{compactitem}
        \item $\Ft.\mathtt{update}(t, n_{tw}/\rbr{n_t+\bar{\beta}})\quad \forall t \in T_{w}$
        \item For each occurrence of $w$, say $w_{i,j} = w$ in $d_i$
        \begin{compactitem}
          \item $t \leftarrow z_{i,j}$ 
          \item Decrease $n_{t}$, $n_{td_i}$, $n_{tw}$ by one  
          \item $\Ft.\mathtt{update}(t, \delta)$ with $\delta = \frac{n_{tw}+\beta}{n_t+\bar{\beta}} - \Ft[\mathtt{leaf}(t)]$
          \item $\cbb \leftarrow \mathtt{cumsum}(\rb)$ (on $T_w$ only)
          \item 
            $t \leftarrow \mathtt{discrete}(\pb,\mathtt{uniform}(\alpha\Ft[1]+\rb^T\one))$
            by \eqref{eq:2levelsampler}
          \item Increase $n_{t}$, $n_{td_i}$, $n_{tw}$ by one  
          \item $\Ft.\mathtt{update}(t, \delta)$ with $\delta = \frac{n_{tw}+\beta}{n_t+\bar{\beta}} - \Ft[\mathtt{leaf}(t)]$
          \item $z_{i,j} \leftarrow t$
        \end{compactitem}
        \item $\Ft.\mathtt{update}(t, -n_{tw}/\rbr{n_t+\bar{\beta}})\quad \forall t \in T_{w}$
      \end{compactitem}
  \end{compactitem}
  \caption{F+LDA with word-by-word sampling}
  \label{alg:F+LDA}
\end{algorithm}

\begin{table*}
  \caption{Comparison of various sampling methods for LDA. We use $\#MH$ to denote number of Metropolis-Hasting steps for Alias LDA. Note that in this table only the order of time complexity is presented---there are some hidden coefficients which also play important roles in practice. For example, the initialization cost of alias table is much slower than linear search although they have the same time complexity. }
  \label{label:}

  \begin{center}
    %\resizebox{18cm}{!}{
    \resizebox{1.1\linewidth}{!}{
    \label{table:spLDA}
    \hspace{-4em}
    \begin{tabular}{l@{ }c@{}c||c@{}c||c@{}c@{}c||c@{}c}
      & \multicolumn{2}{c}{F+LDA} & \multicolumn{2}{c}{F+LDA} & \multicolumn{3}{c}{Sparse-LDA} & \multicolumn{2}{c}{Alias-LDA} \\
      Sample Sequence  & \multicolumn{2}{c}{Word-by-Word} & \multicolumn{2}{c}{Doc-by-Doc} & \multicolumn{3}{c}{Doc-by-Doc} & \multicolumn{2}{c}{Doc-by-Doc} \\
      Exact Sampling    & \multicolumn{2}{c}{Yes} &\multicolumn{2}{c}{Yes} &\multicolumn{3}{c}{Yes}&\multicolumn{2}{c}{No} \\
      \hline
      Decomposition & $\alpha\rbr{\frac{n_{tw}+\beta}{n_t+\bar{\beta}}}$ & $+n_{td}\rbr{\frac{n_{tw}+\beta}{n_t+\bar{\beta}}}$ 
      & $\beta\rbr{\frac{n_{td}+\alpha}{n_t+\bar{\beta}}}$ & $+n_{tw}\rbr{\frac{n_{td}+\alpha}{n_t+\bar{\beta}}}$
      & $\frac{\alpha\beta}{n_t+\bar{\beta}}$ & $+\beta \rbr{\frac{n_{td}}{n_t+\bar{\beta}}}$ & $+n_{tw}\rbr{\frac{n_{td}+\alpha}{n_t+\bar{\beta}}}$ 
      & $\alpha\rbr{\frac{n_{tw}+\beta}{n_t+\bar{\beta}}}$ & $+n_{td}\rbr{\frac{n_{tw}+\beta}{n_t+\bar{\beta}}}$ \\
      Sampling method   & F+tree & BSearch & F+tree & BSearch & LSearch & LSearch & LSearch &  Alias & Alias \\
      Fresh samples     & Yes& Yes& Yes& Yes& Yes& Yes& Yes& No & Yes \\
      Initialization & $\Theta(\log T)$ & $\Theta(\abs{T_d})$ & $\Theta(\log T)$ & $\Theta(\abs{T_w})$ 
      & $\Theta(1)$ & $\Theta(1)$ & $\Theta(\abs{T_w})$ 
      & $\Theta(1)$ & $\Theta(\abs{T_d})$ \\
      Sampling &      $\Theta(\log T)$ & $\Theta(\log \abs{T_d})$ & $\Theta(\log T)$ & $\Theta(\log \abs{T_w})$ 
      & $\Theta(T)$ & $\Theta(\abs{T_d})$ & $\Theta(\abs{T_w})$ 
      & $\Theta(\#MH)$ & $\Theta(\#MH)$ \\

    \end{tabular}
  }
  \end{center}
\end{table*}
\subsection{F+LDA = LDA with F+tree Sampling}
\label{sec:treelda}
In this section, we show the details about what to apply F+tree sampling to CGS 
for LDA. Let us focus on a single CGS step in LDA with the current document id $d_i$, the 
current word $w$, and the current topic assignment $t_{\text{cur}}$. For the 
simplicity of description, we further denote $n_{td} = n_{t,d_i,*}$, 
$n_{tw}=n_{t,w,*}$, and $n_{t} = n_{t,*,*}$ and assume $\alpha_t = \alpha, \forall t$,
$\beta_j = \beta, \forall j$, and $\bar{\beta}=J\times\beta$. The multinomial parameter
$\pb$ of the CGS step in \eqref{eq:cgs-update} can be decomposed into two 
terms as follows.  
\begin{align}
  p_t &= \frac{\rbr{n_{td}+\alpha}\rbr{n_{tw}+\beta}}{n_t + \bar{\beta}},\quad 
  \forall t=1,\ldots,T.  \nonumber \\
  &= \beta\rbr{\frac{n_{td}+\alpha}{n_t+\bar{\beta}}} + 
  n_{tw}\rbr{\frac{n_{td}+\alpha}{n_t+\bar{\beta}}}. \label{eq:treelda-doc}
  %\\ &= \alpha\rbr{\frac{n_{tw}+\beta}{n_t+\bar{\beta}}} + 
  %n_{td}\rbr{\frac{n_{tw}+\beta}{n_t+\bar{\beta}}} \label{eq:treelda-word}.
\end{align}
Let $\qb$ and $\rb$ be two vectors with $q_t = \frac{n_{td}+\alpha}{n_t+\bar{\beta}}$ and $r_t = n_{tw}q_t$. 
Some facts and implications about this decomposition: 
\begin{compactenum}[(a)]
\item $\pb = \beta\qb + \rb$.   \label{eq:prop-sum}
This leads to a simple two-level sampling for $\pb$
\[
\mathtt{discrete}(\pb, u) = 
\begin{cases}
  \mathtt{discrete}(\rb, u) & \text{if } u \le \rb^T\one,  \\
  \mathtt{discrete}(\qb, \frac{u-\rb^T\one}{\beta}) & \text{otherwise},
\end{cases}
\]
where $\one$ is the all-one vector and $\pb^T\one$ denotes the normalization 
constant for $\pb$, and $u=\mathtt{uniform}(\pb^T\one)$. This means that 
sampling for $\pb$ can be very fast if $\qb$ and $\rb$ can be sampled 
efficiently.   
\item $\qb$ is always dense but only two elements will be changed at each CGS 
  step if we follow a document-by-document sampling sequence. \label{eq:prop-q} Note 
  $\qb$ only depends on $n_{td}$. Decrement or increment of  
  a single $n_{td}$ only changes a single element of 
  $\qb$. We propose to apply F+tree sampling for $\qb$ for its logarithmic time
  sampling and maintenance. At the beginning of CGS for LDA, a 
  F+tree $\Ft$ for $\qb$ with $q_t = \frac{\alpha}{n_t+\bar{\beta}}$ is 
  constructed in $\Theta(T)$. When the CGS switches to a new document $d_i$, 
  perform the following updates 
  \[
  \Ft.\mathtt{update}(t, \frac{n_{td}}{n_t+\bar{\beta}})\quad \forall t \in T_d:=\cbr{t:n_{td}\neq 0}.
  \]
  When the CGS finishes the sampling for this document, we can perform 
  $\Ft.\mathtt{update}(t, \frac{-n_{td}}{n_t + \bar{\beta}})\ \forall t \in T_d$.  
  Both updates can be done in $\Theta(\abs{T_d} \log T )$. As $\abs{T_d}$ is 
  upper bounded by the number of words in this document, the amortized 
  sampling cost for each word in the document remains $\Theta(\log T)$.  
   
\item $\rb$ is $T_w$ sparse, where $T_w := \cbr{t: n_{tw}\neq 0}$.  
  \label{eq:prob-r} Unlike $\qb$, all the elements of $\rb$ change when we 
  switch from one word to another word in the same document. Moreover, $\rb$ 
  is only used once to computer $\rb^T\one$ and generate at most one sample. 
  Thus, we propose to use BSearch approach to perform the 
  sampling for $\rb$.  In particular, we only calculate the cumulative sum on 
  nonzero elements in $T_w$. Thus, the initialization cost of BSearch is 
  $\Theta(\abs{T_w})$ and the sampling cost is $\Theta(\log \abs{T_w})$.  
\end{compactenum}
 
  \begin{comment}
F+LDA using document-by-document CGS can be summarized as follows.  
\begin{compactitem}
  \item $\Ft.\mathtt{initialize}(\qb)$, with $q_t = \frac{\alpha}{n_t + \bar{\beta}}$
  \item For each document $d_i$ 
    \begin{compactitem}
      \item $\Ft.\mathtt{update}(t, n_{td}/\rbr{n_t+\bar{\beta}})\quad \forall t \in T_{d_i}$
      \item For each word $w = w_{i,j}$ in $d_i$
      \begin{compactitem}
        \item $t \leftarrow z_{i,j}$
        \item Decrease $n_{t}$, $n_{td}$, $n_{tw}$ by one  
        \item $\Ft.\mathtt{update}(t, \frac{n_{td}+\alpha}{n_t+\bar{\beta}} - \Ft[\mathtt{leaf}(t)])$
        \item Construct the sparse CDF $\cbb$ for $\rb$
        \item $u \leftarrow \mathtt{uniform}(\alpha\Ft[1]+c_{T})$
        \item $t \leftarrow \mathtt{discrete}(\pb, u)$ using using $\Ft$ and $\cbb$ 
        \item Increase $n_{t}$, $n_{td}$, $n_{tw}$ by one  
        \item $\Ft.\mathtt{update}(t, \frac{n_{td}+\alpha}{n_t+\bar{\beta}} - \Ft[\mathtt{leaf}(t)])$
      \end{compactitem}
      \item $\Ft.\mathtt{update}(t, -n_{td}/\rbr{n_t+\bar{\beta}})\quad \forall t \in T_{d_i}$
    \end{compactitem}
\end{compactitem}
\end{comment}

{\bf Word-by-word CGS for LDA.}
Other than the traditional document-by-document CGS for LDA, we can also 
consider CGS with the word-by-word sampling sequence. For this sequence, we 
consider another decomposition of \eqref{eq:treelda-doc} as follows.  
\begin{align}
  p_t &= 
  \alpha\rbr{\frac{n_{tw}+\beta}{n_t+\bar{\beta}}} + 
  n_{td}\rbr{\frac{n_{tw}+\beta}{n_t+\bar{\beta}}} \label{eq:treelda-word}, 
  \quad \forall t.
\end{align}
For this decomposition \eqref{eq:treelda-word}, $\qb$ and $\rb$ have analogue 
definitions such that $q_t = \frac{n_{tw}+\beta}{n_{t}+\bar{\beta}}$ and 
$r_t = n_{td} q_t$, respectively. The corresponding three facts for 
\eqref{eq:treelda-word} are as follows.  
\begin{compactenum}[(a)]
\item $\pb = \alpha \qb + \rb$.  The two-level sampling for $\pb$ is 
\begin{equation}
  \mathtt{discrete}(\pb, u) = 
  \begin{cases}
    \mathtt{discrete}(\rb, u) & \text{if } u \le \rb^T\one,  \\
    \mathtt{discrete}(\qb, \frac{u-\rb^T\one}{\alpha}) & \text{otherwise},
  \end{cases}
  \label{eq:2levelsampler}
\end{equation}
\item $\qb$ is always dense but only very few elements will be changed at each 
  CGS step using word-by-word sampling sequence. A F+tree structure $\Ft$ is 
  maintained for $\qb$. The amortized update time for each occurrence of a 
  word is $\Theta(\log T)$ and the sampling generation for $\qb$ using $\Ft$ 
  also costs $\Theta(\log T)$. Thus, 
  $\mathtt{discrete}(\qb, u):= \Ft.\mathtt{sample}(u)$.
\item $\rb$ is a sparse vector with $\abs{T_d}$ non-zeros. BSearch is
  used to constructed $\cbb = \mathtt{cumsum}(\rb)$ in $\Theta(T_d)$ space and 
  time.  $\cbb$ is used to perform binary search to generate a sample required 
  by CGS for the occurrence of the current word. Thus, 
  $\mathtt{discrete}(\rb, u):= \mathtt{binary\_search}(\cbb, u)$.
\end{compactenum}
The detailed procedure of using word-by-word sampling sequence is
described in Algorithm \ref{alg:F+LDA}.  Let us analyse the performance
difference of F+LDA between two sampling sequences of a large number of
documents.  The amortized cost for each CGS step is
$\Theta(|T_d|+\log T)$ for the word-by-word sequence and
$\Theta(|T_w|+\log T)$ for the document-by-document sequence.  Note that
$\abs{T_d}$ is always bounded by the number of words in a document,
which is usually a much smaller number than a large $T$ (say 1024).  In
contrast, $\abs{T_w}$ approaches to $T$ when the number of documents
increases. As a result, we can expect that F+LDA with the word-by-word
sequence has faster performance than the document-by-document sequence.
Empirical results in Section \ref{sec:exp_sampling} also conform our
analysis.

\subsection{Related Work}
SparseLDA~\cite{YaoMimMcC09} is the first sampling method which
considered 
decomposing $\pb$ into a sum of sparse vectors and a dense vector. In 
particular, it considers a three-term decomposition of $p_t$ as follows. 
\begin{align*}
  p_t &= \frac{\alpha\beta}{n_t+\bar{\beta}} +\beta 
  \rbr{\frac{n_{td}}{n_t+\bar{\beta}}} 
  +n_{tw}\rbr{\frac{n_{td}+\alpha}{n_t+\bar{\beta}}}, 
\end{align*}
where the first term is dense, the second term is sparse with $\abs{T_d}$ 
non-zeros, and the third term is sparse with $\abs{T_w}$. 
In both SparseLDA implementations (Yahoo!~LDA \cite{SmoNar10} and Mallet LDA 
\cite{YaoMimMcC09}), LSearch is applied for all of these three terms. As 
SparseLDA follows the document-by-document sequence, only very few elements 
will be changed for the first two terms at each CGS step. Sampling procedures 
for the first two term have very low chance to be performed due to the 
observation that most mass of $p_t$ is contributed from the third term. The 
choice of LSearch, whose normalization constant $c_T$ can be updated in 
$\Theta(1)$, for the first two term is reasonable. Note that $\Theta(T)$ and 
$\Theta(\abs{T_d})$ initialization costs for the first two term  can be 
amortized. The overall amortized cost for each CGS step is 
$\Theta(|T_w|+|T_d|+|T|)$.  

AliasLDA~\cite{LiAhmRavSmo14} is a recent proposed approach which reduces the 
amortized cost of each step to $\Theta(|T_d|)$. AliasLDA considers the 
following decomposition on $\pb$: 
\begin{align*}
  p_t &=\alpha\rbr{\frac{n_{tw}+\beta}{n_t+\bar{\beta}}}+n_{td}\rbr{\frac{n_{tw}+\beta}{n_t+\bar{\beta}}}.
\end{align*}
In stead of the ``exact'' multinomial sampling for $\pb$, AliasLDA considers a 
proposal distribution $\qb$ with a very efficient generate routine and perform 
a series of Metropolis-Hasting (MH) steps using this proposal to simulate the true 
distribution $\pb$. In particular, the proposal distribution is 
constructed using the latest second term and a stale version of the first
term. For both terms, Alias method is applied to perform the sampling. $\#MH$ 
steps decides the quality of the sampling results. The overall amortized 
cost for each CGS step is $\Theta(|T_d|+\#MH)$. Note the initialization cost 
$\Theta(|T|)$ for the first term can be amortized as long as the same Alias 
table can be used to generate $T$ samples.  

See Table \ref{table:spLDA} for a detailed summary for LDA using various 
sampling methods. Note that the hidden coefficient $\rho_A$ in the $\Theta(|T_d|)$ notation 
for the construction of the Alias table is larger than the coefficient $\rho_B$ 
for the construction of BSearch and the coefficient $\rho_F$ for the 
maintenance and sampling of F+tree. Thus as long as 
$T < 2^{\frac{\rho_A-\rho_B}{\rho_F}|T_d|}$, F+LDA using the word-by-word 
sampling sequence is faster than AliasLDA. Empirical results in Section 
\ref{sec:exp_sampling} also shows the superiority of F+LDA over AliasLDA for 
real-world datasets using $T=1024$.
 
\section{Proposed Parallel Approach}

\begin{figure}[t]
  \begin{center}
    % \begin{subfigure}[t]{0.3\textwidth}
    %   \centering      
    %   \begin{tikzpicture}[scale=0.5]
    %     \draw[help lines,white] (0,0) grid (6,8);
    %     \node at (0.5, -0.6) {words};
    %     \node at (4.5, -0.6) {documents};
    %       %     \node[gnode] (root) at (2.5, 7) {};
    %       %     \node at (3.5,7) {$\sbb$};
    %       %     \node at (2.5,8.5) {summation node};
    
    %     \node at (5.5,3.5) {$\db_i$};
    %     \node[gnode] (u1) at (4.5,5.5) {};
    %     \node[gnode] (u2) at (4.5,4.5) {};
    %     \node[gnode] (u3) at (4.5,3.5) {};
    %     \node[gnode] (u4) at (4.5,2.5) {};
    %     \node[gnode] (u5) at (4.5,1.5) {};
    %     \node[gnode] (u6) at (4.5,0.5) {};
    
    %     \node at (-.5,4.5) {$\wb_j$};
    %     \node[gnode] (m1) at (0.5,4.5) {};
    %     \node[gnode] (m2) at (0.5,3.5) {};
    %     \node[gnode] (m3) at (0.5,2.5) {};
    %     \node[gnode] (m4) at (0.5,1.5) {};
    
    %     \draw[-,thick,gray] (u1) -- (m1);
    %     \draw[-,thick,gray] (u1) -- (m4);
    
    %     \draw[-,thick,gray] (u2) -- (m2);
    
    %     %     \draw[-,ultra thick,black] (u3) -- (m1);
    %     \draw[-,thick,gray] (u3) -- (m1);
    %     \draw[-,thick,gray] (u3) -- (m3);
    
    %     \draw[-,thick,gray] (u4) -- (m2);
    
    %     \draw[-,thick,gray] (u5) -- (m1);
    %     \draw[-,thick,gray] (u5) -- (m4);
    
    %     \draw[-,thick,gray] (u6) -- (m3);
    
    %   \end{tikzpicture}
    %   \caption{Graph for document corpus\label{fig:corpus-graph}}
    % \end{subfigure}
    % \quad
    \begin{subfigure}[t]{0.22\textwidth}
      \centering      
      \begin{tikzpicture}[scale=0.5]
        \draw[help lines,white] (0,0) grid (6,8);
        \node at (0.5, -0.6) {words};
        \node at (4.5, -0.6) {documents};
        \node[rnode] (root) at (2.5, 7) {};
        \node at (3.5,7) {$\sbb$};
        \node at (2.5,8.5) {summation node};
        
        \node at (5.5,3.5) {$\db_i$};
        \node[gnode] (u1) at (4.5,5.5) {};
        \node[gnode] (u2) at (4.5,4.5) {};
        \node[rnode] (u3) at (4.5,3.5) {};
        \node[gnode] (u4) at (4.5,2.5) {};
        \node[gnode] (u5) at (4.5,1.5) {};
        \node[gnode] (u6) at (4.5,0.5) {};
        
        \node at (-.5,4.5) {$\wb_j$};
        \node[rnode] (m1) at (0.5,4.5) {};
        \node[gnode] (m2) at (0.5,3.5) {};
        \node[gnode] (m3) at (0.5,2.5) {};
        \node[gnode] (m4) at (0.5,1.5) {};
        
        \draw[-,ultra thick,black] (root) -- (u3);
        \draw[-,ultra thick,black] (root) -- (m1);
        
        \draw[-,thick,gray] (u1) -- (m1);
        \draw[-,thick,gray] (u1) -- (m4);
        
        \draw[-,thick,gray] (u2) -- (m2);
        
        \draw[-,ultra thick,black] (u3) -- (m1);
        \draw[-,thick,gray] (u3) -- (m3);
        
        \draw[-,thick,gray] (u4) -- (m2);
        
        \draw[-,thick,gray] (u5) -- (m1);
        \draw[-,thick,gray] (u5) -- (m4);
        
        \draw[-,thick,gray] (u6) -- (m3);
        
        % \node at (-0.5,3.0) {(b)};
      \end{tikzpicture}
      \caption{Access graph for LDA\label{fig:lda-graph}}
    \end{subfigure}
    \quad
    \begin{subfigure}[t]{0.22\textwidth}
      \centering      
      \begin{tikzpicture}[scale=1.0]
        
        % \fill [black!20!white] (2.4,2) rectangle (2.6,3);
        % \fill [black!20!white] (0.2,1) rectangle (0.4,2);
        \fill [black!20!white] (0.1875,0.75) rectangle (0.375,1.5);
        
        \begin{scope}[scale=0.375]
          \dtpt{5}{1}  \dtpt{0}{15}  \dtpt{13}{12}  \dtpt{13}{6}  \dtpt{14}{8}  \dtpt{0}{8}  \dtpt{7}{10}  \dtpt{15}{2}  \dtpt{10}{5}  \dtpt{8}{13}  \dtpt{5}{3}  \dtpt{12}{5}  \dtpt{10}{7}  \dtpt{11}{6}  \dtpt{15}{3}  \dtpt{6}{9}  \dtpt{13}{5}  \dtpt{1}{3}  \dtpt{8}{15}  \dtpt{1}{6}  \dtpt{1}{7}  \dtpt{8}{9}  \dtpt{12}{4}  \dtpt{12}{1}  \dtpt{13}{10}  \dtpt{9}{0}  \dtpt{8}{4}  \dtpt{15}{12}  \dtpt{3}{6}  \dtpt{10}{13} 
        \end{scope}
        
        % \draw [xstep=0.2, ystep=1.0] (0,0) grid (3,3);
        \draw [xstep=0.1875, ystep=0.75] (0,0) grid (3,3);
        % \draw [vthick] (0,0) rectangle (3,3);
        \draw [very thick] (0,2.25) rectangle (3,3);
        \draw [very thick] (0,1.5) rectangle (3,2.25);
        \draw [very thick] (0,0.75) rectangle (3,1.5);
        \draw [very thick] (0,0) rectangle (3,0.75);
      \end{tikzpicture}
      \caption{Task and Data Partition\label{fig:nomad-split}}
    \end{subfigure}
  \end{center}
  \caption{Abstract access graph for LDA}
  \label{fig:lda-graph}
\end{figure}
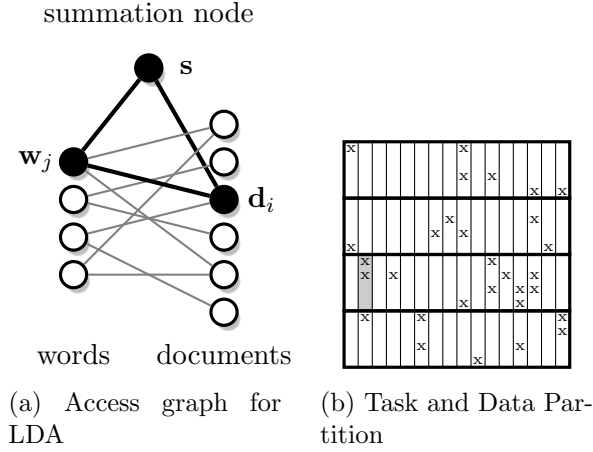

% To explain our proposed approach we find it instructive to consider a
% graph $G=\rbr{V, E}$ with $\rbr{I + J + 1}$ nodes. The documents $d_{i}$
% for $i=1,\ldots,I$ correspond to $I$ nodes and the words $w_{j}$ for
% $j=1,\ldots,J$ correspond to the remaining $J$ nodes. We add an edge
% whenever word $w_{j}$ occurs in document $d_{i}$. Moreover, there is one
% global node which is connected to all documents and all words. See
% Figure \ref{fig:lda-graph} (a) for a visual depiction. 
In this section we present our second innovation---a novel parallel
framework for CGS. 
%various types of inference methods for topic modeling.
%We apply this parallel framework to Collapsed Gibbs
%sampling in this paper, but 
Note that the same technique can also be used for other inference
techniques for LDA such as collapsed variational Bayes
and stochastic variational Bayes~\cite{AsuWelSmyTeh09}
since they follow the similar update pattern. 

To explain our proposed approach, we find it instructive to consider a
hyper graph $G$. Let $G = (V,E)$ be a hyper graph with $(I+J+1)$
nodes:
\[
V = \{\db_i: i=1,\dots,I\} \cup \{\wb_j: j = 1,\dots,J\} \cup \{\sbb\},
\]
and hyper edges: 
\[
E = \{e_{ij} = \{\db_i, \wb_{j}, \sbb\}\},
\]
where $|E| = \sum_i n_i$.  Note that $G$ contains multi-edges, which
means that the same hyper edge can appear more than once in $E$ just
as a single word can appear multiple times in a document.  Clearly,
$G$ is equivalent to a bag-of-the-words representation of the corpus
$\{d_1,\dots,d_I\}$; each $\db_i$ is associated with the $i$-th
document, each $\wb_j$ is associated with the $j$-th vocabulary, and
each hyper edge $e_{ij}$ corresponds to one occurrence of the
vocabulary $w_{j}$ in the $i$-th document $d_i$.  See Figure
\ref{fig:lda-graph} (a) for a visual illustration; here, each gray
edge corresponds to an occurrence of a word and the black triangle
highlights a particular hyper edge $e_{ij}=\cbr{\db_i, \wb_j, \sbb}$.

\par
To further connect $G$ to the update rule of CGS, 
we associate each node of $G$ with a $T$-dimensional vector.
In many inference methods, an update based on a single occurrence
$w_{ij}$ can be realized as a graph operation on $G$ which accesses
values of nodes in a single hyper edge $e_{ij}$.  More concretely, let
us define the $t$-th coordinate of each vector as follows:
\[
(\db_i)_{t} := n_{t,i,*},\quad (\wb_j)_t := n_{t,*,w_j}, \text{ and}\quad 
(\sbb)_{t} := n_{t,*,*}. 
\]
Based on the update rule of CGS, we can see that the update for
the occurrence of $w_{ij}$ only reads from and writes to the values
stored in $\db_i$, $\wb_{w_{ij}}$, and $\sbb$.

Interestingly, this property of the updates is reminiscent of that of
the stochastic gradient descent (SGD) algorithm for matrix completion
model.  Similarly to LDA, matrix completion model has two sets of
parameters $\wb_1,\ldots,\wb_J$ and $\db_1,\ldots,\db_I$, and each SGD
update requires only one of $\wb_j$ and one of $\db_i$ to be read and
modified.  Since each update is highly localized, there is a huge room
for parallelization; \cite{YunYuHsietal13} exploit this property to
propose an efficient asynchronous parallel SGD algorithm for matrix
completion.

The crucial difference in the case of LDA, however, is that there is an
additional variable $\sbb$ which participates in every hyper edge of the
graph.  Therefore, if we change the update sequence from $(e_{ij},
e_{i'j'})$ to $(e_{i'j'}, e_{ij})$, then even if $i \neq i'$ and $j \neq
j'$ the result of updates will not be the same since the value of $\sbb$
changes in the first update.  Fortunately, this dependency is very weak;
since each element $\sbb$ is a large number because it is a summation
over the whole corpus and each update can change its value at most by
one, the relative change of $\sbb$ made in a short period of time is often
negligible.

While existing approaches such as Yahoo!\ LDA \cite{SmoNar10} exploit
this observation by introducing a parameter server and let each machine
to query the server to retrieve recent updates, it is certainly not
desirable in the large scale systems that every machine has to query the
same central server.  Motivated by the ``nomadic'' algorithm introduced
by \cite{YunYuHsietal13} for matrix completion, we propose a new
parallel framework for LDA that is decentralized, asynchronous and
lock-free.

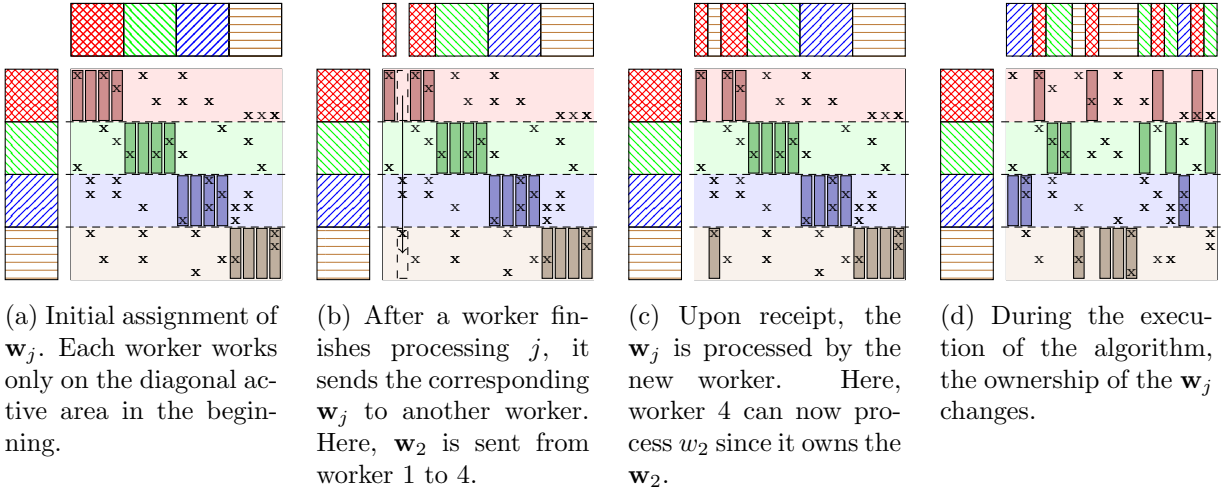
\begin{figure}[htb]
  %\hspace{-2em}
  \centering
  \begin{subfigure}[t]{0.22\textwidth}
    \centering
    \begin{tikzpicture}[scale=0.35]
      
      \draw (0,0) rectangle (\gridwidth * \nodenum, \gridwidth * \nodenum);
      
      \fill [red!10] (0, 3 * \gridwidth) 
      rectangle (\gridwidth * \nodenum, 4 * \gridwidth);

      \fill [green!10] (0, 2 * \gridwidth) 
      rectangle (\gridwidth * \nodenum, 3 * \gridwidth);

      \fill [blue!10] (0, 1 * \gridwidth) 
      rectangle (\gridwidth * \nodenum, 2 * \gridwidth);

      \fill [brown!10] (0,0) 
      rectangle (\gridwidth * \nodenum, 1 * \gridwidth);

      % split between nodes
      \foreach \y in {1,2,3}
      {
        \draw[densely dashed] 
        (-\gridwidth * 0.1, \gridwidth * \y)
        --
        (\gridwidth * \nodenum + \gridwidth * 0.1,  \gridwidth * \y);
      }
      
      \drawrowpart

      \drawcolpart{0}{3}{\ptrna}{\ptcla}
      \drawcolpart{4}{7}{\ptrnb}{\ptclb}
      \drawcolpart{8}{11}{\ptrnc}{\ptclc}
      \drawcolpart{12}{15}{\ptrnd}{\ptcld}

      \dtpt{5}{1} \dtpt{0}{15} \dtpt{13}{12} \dtpt{13}{6} \dtpt{14}{8}
      \dtpt{0}{8} \dtpt{7}{10} \dtpt{15}{2} \dtpt{10}{5} \dtpt{8}{13}
      \dtpt{5}{3} \dtpt{12}{5} \dtpt{10}{7} \dtpt{11}{6} \dtpt{15}{3}
      \dtpt{6}{9} \dtpt{13}{5} \dtpt{1}{3} \dtpt{8}{15} \dtpt{1}{6}
      \dtpt{1}{7} \dtpt{8}{9} \dtpt{12}{4} \dtpt{12}{1} \dtpt{13}{10}
      \dtpt{9}{0} \dtpt{8}{4} \dtpt{15}{12} \dtpt{3}{6} \dtpt{10}{13}
      \dtpt{6}{13} \dtpt{2}{11} \dtpt{5}{5} \dtpt{2}{1} \dtpt{3}{7}
      \dtpt{9}{3} \dtpt{11}{1} \dtpt{3}{14} \dtpt{11}{11} \dtpt{5}{15}

      \drawcolpart{0}{3}{\ptrna}{\ptcla}
      \drawcolpart{4}{7}{\ptrnb}{\ptclb}
      \drawcolpart{8}{11}{\ptrnc}{\ptclc}
      \drawcolpart{12}{15}{\ptrnd}{\ptcld}

      \asnbox{0}{0}{red} \asnbox{1}{0}{red} \asnbox{2}{0}{red} \asnbox{3}{0}{red}
      \asnbox{4}{1}{green} \asnbox{5}{1}{green} \asnbox{6}{1}{green} \asnbox{7}{1}{green}
      \asnbox{8}{2}{blue} \asnbox{9}{2}{blue} \asnbox{10}{2}{blue} \asnbox{11}{2}{blue}
      \asnbox{12}{3}{brown} \asnbox{13}{3}{brown} \asnbox{14}{3}{brown} \asnbox{15}{3}{brown}
      
      \dtpt{5}{1} \dtpt{0}{15} \dtpt{13}{12} \dtpt{13}{6} \dtpt{14}{8}
      \dtpt{0}{8} \dtpt{7}{10} \dtpt{15}{2} \dtpt{10}{5} \dtpt{8}{13}
      \dtpt{5}{3} \dtpt{12}{5} \dtpt{10}{7} \dtpt{11}{6} \dtpt{15}{3}
      \dtpt{6}{9} \dtpt{13}{5} \dtpt{1}{3} \dtpt{8}{15} \dtpt{1}{6}
      \dtpt{1}{7} \dtpt{8}{9} \dtpt{12}{4} \dtpt{12}{1} \dtpt{13}{10}
      \dtpt{9}{0} \dtpt{8}{4} \dtpt{15}{12} \dtpt{3}{6} \dtpt{10}{13}
      \dtpt{6}{13} \dtpt{2}{11} \dtpt{5}{5} \dtpt{2}{1} \dtpt{3}{7}
      \dtpt{9}{3} \dtpt{11}{1} \dtpt{3}{14} \dtpt{11}{11} \dtpt{5}{15}

      \dtpt{2}{15} \dtpt{3}{10} \dtpt{4}{9} \dtpt{15}{12} \dtpt{14}{12}

    \end{tikzpicture}
    \caption{Initial assignment of $\wb_j$. Each worker works
      only on the diagonal active area in the beginning.}
  \end{subfigure}
  \quad
  \begin{subfigure}[t]{0.22\textwidth}
    \centering
    \begin{tikzpicture}[scale=0.35]

      \boxlayout
      
      \drawrowpart

      \drawcolpart{0}{0}{\ptrna}{\ptcla}
      \drawcolpart{2}{3}{\ptrna}{\ptcla}
      \drawcolpart{4}{7}{\ptrnb}{\ptclb}
      \drawcolpart{8}{11}{\ptrnc}{\ptclc}
      \drawcolpart{12}{15}{\ptrnd}{\ptcld}

      \asnbox{0}{0}{red} \asnbox{2}{0}{red} \asnbox{3}{0}{red}
      \asnbox{4}{1}{green} \asnbox{5}{1}{green} \asnbox{6}{1}{green} \asnbox{7}{1}{green}
      \asnbox{8}{2}{blue} \asnbox{9}{2}{blue} \asnbox{10}{2}{blue} \asnbox{11}{2}{blue}
      \asnbox{12}{3}{brown} \asnbox{13}{3}{brown} \asnbox{14}{3}{brown} \asnbox{15}{3}{brown}
      
      \dasnbox{1}{0} \dasnbox{1}{3}
      \mvarr{1}{0}{1}{3}

      \dtpt{5}{1} \dtpt{0}{15} \dtpt{13}{12} \dtpt{13}{6} \dtpt{14}{8}
      \dtpt{0}{8} \dtpt{7}{10} \dtpt{15}{2} \dtpt{10}{5} \dtpt{8}{13}
      \dtpt{5}{3} \dtpt{12}{5} \dtpt{10}{7} \dtpt{11}{6} \dtpt{15}{3}
      \dtpt{6}{9} \dtpt{13}{5} \dtpt{1}{3} \dtpt{8}{15} \dtpt{1}{6}
      \dtpt{1}{7} \dtpt{8}{9} \dtpt{12}{4} \dtpt{12}{1} \dtpt{13}{10}
      \dtpt{9}{0} \dtpt{8}{4} \dtpt{15}{12} \dtpt{3}{6} \dtpt{10}{13}
      \dtpt{6}{13} \dtpt{2}{11} \dtpt{5}{5} \dtpt{2}{1} \dtpt{3}{7}
      \dtpt{9}{3} \dtpt{11}{1} \dtpt{3}{14} \dtpt{11}{11} \dtpt{5}{15}
      \dtpt{2}{15} \dtpt{3}{10} \dtpt{4}{9} \dtpt{15}{12} \dtpt{14}{12}

    \end{tikzpicture}
    \caption{After a worker finishes processing $j$, it sends
      the corresponding $\wb_j$ to another worker.
      Here, $\wb_2$ is sent from worker $1$ to $4$.  }
  \end{subfigure}
  \quad
  %\hspace{2em}
  \begin{subfigure}[t]{0.22\textwidth}
    \centering
    \begin{tikzpicture}[scale=0.35]

      \boxlayout
      
      \drawrowpart

      \drawcolpart{0}{0}{\ptrna}{\ptcla}
      \drawcolpart{1}{1}{\ptrnd}{\ptcld}
      \drawcolpart{2}{3}{\ptrna}{\ptcla}
      \drawcolpart{4}{7}{\ptrnb}{\ptclb}
      \drawcolpart{8}{11}{\ptrnc}{\ptclc}
      \drawcolpart{12}{15}{\ptrnd}{\ptcld}

      \asnbox{0}{0}{red} \asnbox{2}{0}{red} \asnbox{3}{0}{red}
      \asnbox{4}{1}{green} \asnbox{5}{1}{green} \asnbox{6}{1}{green} \asnbox{7}{1}{green}
      \asnbox{8}{2}{blue} \asnbox{9}{2}{blue} \asnbox{10}{2}{blue} \asnbox{11}{2}{blue}
      \asnbox{12}{3}{brown} \asnbox{13}{3}{brown} \asnbox{14}{3}{brown} \asnbox{15}{3}{brown}
      \asnbox{1}{3}{brown}

      \dtpt{5}{1} \dtpt{0}{15} \dtpt{13}{12} \dtpt{13}{6} \dtpt{14}{8}
      \dtpt{0}{8} \dtpt{7}{10} \dtpt{15}{2} \dtpt{10}{5} \dtpt{8}{13}
      \dtpt{5}{3} \dtpt{12}{5} \dtpt{10}{7} \dtpt{11}{6} \dtpt{15}{3}
      \dtpt{6}{9} \dtpt{13}{5} \dtpt{1}{3} \dtpt{8}{15} \dtpt{1}{6}
      \dtpt{1}{7} \dtpt{8}{9} \dtpt{12}{4} \dtpt{12}{1} \dtpt{13}{10}
      \dtpt{9}{0} \dtpt{8}{4} \dtpt{15}{12} \dtpt{3}{6} \dtpt{10}{13}
      \dtpt{6}{13} \dtpt{2}{11} \dtpt{5}{5} \dtpt{2}{1} \dtpt{3}{7}
      \dtpt{9}{3} \dtpt{11}{1} \dtpt{3}{14} \dtpt{11}{11} \dtpt{5}{15}
      \dtpt{2}{15} \dtpt{3}{10} \dtpt{4}{9} \dtpt{15}{12}
      \dtpt{14}{12}

    \end{tikzpicture}
    \caption{Upon receipt, the $\wb_j$ is processed by the new worker.
      Here, worker $4$ can now process $w_2$ since it owns the
      $\wb_2$.}
  \end{subfigure}
  \quad
  \begin{subfigure}[t]{0.22\textwidth}
    \centering
    \begin{tikzpicture}[scale=0.35]

      \boxlayout

      \drawrowpart

      \asnbox{0}{2}{blue} \asnbox{1}{2}{blue} \asnbox{2}{0}{red} \asnbox{3}{1}{green}
      \asnbox{4}{1}{green} \asnbox{5}{3}{brown} \asnbox{6}{0}{red} \asnbox{7}{3}{brown}
      \asnbox{8}{3}{brown} \asnbox{9}{3}{brown} \asnbox{10}{1}{green} \asnbox{11}{0}{red}
      \asnbox{12}{1}{green} \asnbox{13}{2}{blue} \asnbox{14}{0}{red} \asnbox{15}{1}{green}

      \drawcolpart{0}{1}{\ptrnc}{\ptclc}
      \drawcolpart{2}{2}{\ptrna}{\ptcla}
      \drawcolpart{3}{4}{\ptrnb}{\ptclb}
      \drawcolpart{5}{5}{\ptrnd}{\ptcld}
      \drawcolpart{6}{6}{\ptrna}{\ptcla}
      \drawcolpart{7}{9}{\ptrnd}{\ptcld}
      \drawcolpart{10}{10}{\ptrnb}{\ptclb}
      \drawcolpart{11}{11}{\ptrna}{\ptcla}
      \drawcolpart{12}{12}{\ptrnb}{\ptclb}
      \drawcolpart{13}{13}{\ptrnc}{\ptclc}
      \drawcolpart{14}{14}{\ptrna}{\ptcla}
      \drawcolpart{15}{15}{\ptrnb}{\ptclb}
      
      \dtpt{5}{1} \dtpt{0}{15} \dtpt{13}{12} \dtpt{13}{6} \dtpt{14}{8}
      \dtpt{0}{8} \dtpt{7}{10} \dtpt{15}{2} \dtpt{10}{5} \dtpt{8}{13}
      \dtpt{5}{3} \dtpt{12}{5} \dtpt{10}{7} \dtpt{11}{6} \dtpt{15}{3}
      \dtpt{6}{9} \dtpt{13}{5} \dtpt{1}{3} \dtpt{8}{15} \dtpt{1}{6}
      \dtpt{1}{7} \dtpt{8}{9} \dtpt{12}{4} \dtpt{12}{1} \dtpt{13}{10}
      \dtpt{9}{0} \dtpt{8}{4} \dtpt{15}{12} \dtpt{3}{6} \dtpt{10}{13}
      \dtpt{6}{13} \dtpt{2}{11} \dtpt{5}{5} \dtpt{2}{1} \dtpt{3}{7}
      \dtpt{9}{3} \dtpt{11}{1} \dtpt{3}{14} \dtpt{11}{11} \dtpt{5}{15}

      \dtpt{2}{15} \dtpt{3}{10} \dtpt{4}{9} \dtpt{15}{12} \dtpt{14}{12}

    \end{tikzpicture}
    \caption{During the execution of the algorithm, the ownership of the
      $\wb_j$ changes.}
  \end{subfigure}
  \caption{Illustration of the Nomad LDA algorithm}
  \label{fig:nomad_scheme}
\end{figure}

\subsection{Nomadic Framework for Parallel LDA}
Let $p$ be the number of parallel workers, which can be a thread in a
shared-memory multi-core machine or a processor in a distributed memory 
multi-machine system. 

\par
{\bf Data Partition and Subtask Split.} The given document corpus is split into 
$p$ portions such that the $l$-th worker owns the $l$-th partition of the 
data, $D_l \subset \{1,\dots,J\}$. 
Unlike the other parallel approach where each unit subtask is a document owned by the 
worker, our approach uses a fine-grained split for tasks. Note that in 
the inference for LDA, each word occurrence corresponds to a update. Thus, we 
consider a unit subtask $\tb_j$ as all occurrence of word $w_j$ in all documents owned 
by the worker. See Figure \ref{fig:nomad-split} for an illustration on the 
data partition and task split. Each ``x'' denotes an occurrence of a word. 
Each block row (bigger rectangle) represents a  
data partition owned by a worker, while each smaller rectangle stands for a 
unit subtask for the worker.

\par
{\bf Asynchronous Computation.} 
It is known that synchronous computation would suffer from {\em the
  curse of last reducer} when the load-balance is poor.  In this work,
we aim to develop an asynchronous parallel framework where each worker
maintains a local job queue $\qb_l$ such that the worker can keep
performing the subtask popped from the queue without worrying about data
conflict and synchronization. To achieve this goal, we first study the
characteristics of subtasks.  The subtask $\tb_j$ for the $l$-th worker
involves the updates on the all occurrences of $\wb_j$ in $D_l$, which
means that to perform $\tb_j$, the $l$-th worker must acquire the
permission to access $\{\db_i: i \in D_l\}$, $\wb_j$, and $\sbb$. Our
data partition scheme has guaranteed that two workers will never need to
access a same $\db_i$ simultaneously. Thus we can always keep the
ownership of $\db_i, \forall i \in D_l$ to $l$-th worker. The difficulty
for parallel execution comes from the access to $\wb_j$ and $\sbb$ which
can be accessed by different workers at the same time.  To overcome this
difficulty, we propose to use a {\em nomadic token passing} scheme to
avoid access conflicts.  Token passing is a standard technique used in
telecommunication to avoid conflicting access to a resource shared by
many members. The idea is ``owner computes:'' only the member with the
ownership of the token has the permission to access the shared resource.
Here we borrow the same idea to avoid the situation where two workers
require access to the same $\wb_j$ and $\sbb$.

\par
{\bf Nomadic Tokens for $\wb_j$.}
We have a word token $\tau_j$ dedicated for the ownership for each $\wb_j$. 
These $J$ tokens are nomadically passed among $p$ workers. The ownership of a 
token  $\tau_j$ means the worker can perform the subtask $\tb_j$.  Each token 
$\tau_j$ is a tuple $(j, \wb_j)$, where the first entry is the index for the 
token, and the second entry is the latest values of $\wb_j$. For a worker, a 
token $\tau$ means the activation of the corresponding inference subtask. As 
a result, we can guarantee that 1) the values of $\wb_j$ used in each subtask is 
always up-to-date; 2) no two workers require access to a same $\wb_j$. 

\par
{\bf Nomadic Token for $\sbb$.} 
So far we have successfully keep the values of $\db_i$ and $\wb_j$ used in each
subtask latest and avoid the conflicting access by nomadic token passing. 
However, the property which all updates require the access of $\sbb$ makes all 
subtasks depend on each other. Based on the summation property, we proposed to 
deal with this issue by a special nomadic token $\tau_s = (0, \sbb)$ for $\sbb$, 
where $0$ is the token index for $\tau_s$, and two copies of $\sbb$ in each worker: $\sbb_l$ and $\sbbar$.
$\sbb_l$ is a local shadow node for $\sbb$. The $l$-th worker always uses the  
values of $\sbb_l$ to perform updates and makes the modification to $\sbb_l$. 
$\sbbar$ was the snapshot of $\sbb$ from the last arrival of $\tau_s$. Due to 
the additivity of $\sbb$, the delta $\sbb_l - \sbb$ can be regarded as 
the effort that has been made since the last arrival of $\tau_s$. Thus, 
each time when the $\tau_s$ arrives, the worker can performs the following 
operations to accumulate its local effort to the global $\sbb$ and update its 
local $\sbb_l$. 
\begin{enumerate}
\item $\sbb \leftarrow \sbb + \left(\sbb_l-\sbbar\right)$
\item $\sbbar \leftarrow \sbb$
\item $\sbb_l \leftarrow \sbb$
\end{enumerate}

% \subsection{Detailed Updates Rules for Collapsed Gibbs Sampling}
We then present the general idea of Nomad LDA in Algorithm 
\ref{alg:nomad} and an illustration in Figure \ref{fig:nomad_scheme}. 

% \begin{itemize}
% \item TBB concurrent queue
% \item MPI
% \item accelerated sampling techniques for CGS
% \end{itemize}
\begin{algorithm}[htbp]
  \begin{itemize}
  \item[] Given: initialized $\sbb_l$, $\sbbar$, and local queue $\qb_l$
  \item {\bf While} stop signal has not been received
    \begin{itemize}
    \item If receive a token $\tau$, $push(\qb_l, \tau)$
    \item $\tau \leftarrow pop(\qb_l)$ 
    \item If $\tau = \tau_s$
      \begin{itemize}
      \item  $\sbb \leftarrow \sbb + (\sbb_l - \sbbar)$
      \item $\sbb_l \leftarrow \sbb$
      \item $\sbbar \leftarrow \sbb$
      \item Send $\tau_s$ to another worker
      \end{itemize}
    \item Else if $\tau = \tau_j := (j, \wb_k)$
      \begin{itemize}
      \item Perform the $j$-th subtask
      \item Send $\tau_s$ to another worker
      \end{itemize}
    \end{itemize}
  \end{itemize}
  \caption{The basic Nomad LDA algorithm}
  \label{alg:nomad}
\end{algorithm}
\begin{figure*}[tbh]
  \begin{center}
  \begin{tabular}{cc}
  \begin{subfigure}[t]{0.4\textwidth}
    \includegraphics[width=\linewidth]{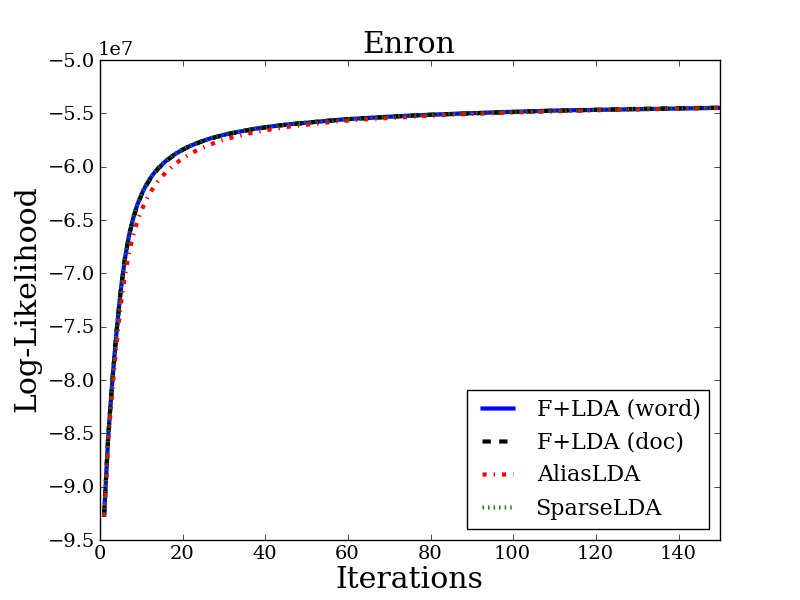}
    \caption{}
    \label{fig:enron-conv}
  \end{subfigure}
  &
  \begin{subfigure}[t]{0.4\textwidth}
    \includegraphics[width=\linewidth]{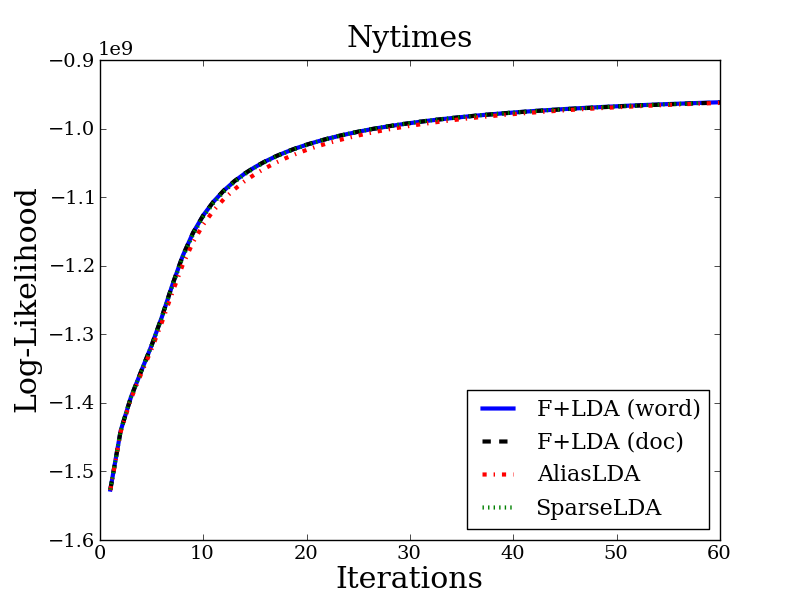}
    \caption{}
    \label{fig:nytimes-conv}
  \end{subfigure}
  \\
  \begin{subfigure}[t]{0.4\textwidth}
    \includegraphics[width=\linewidth]{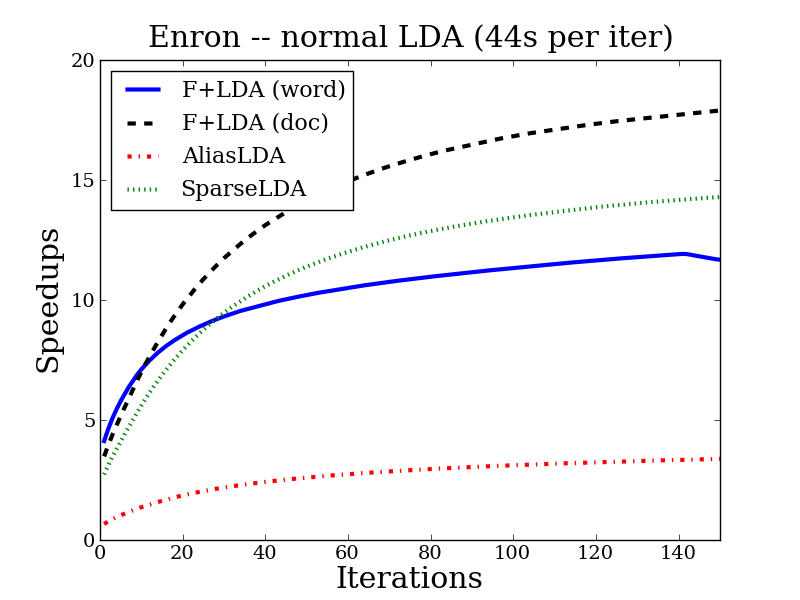}
    \caption{}
    \label{fig:enron-speed}
  \end{subfigure}
  &
  \begin{subfigure}[t]{0.4\textwidth}
    \includegraphics[width=\linewidth]{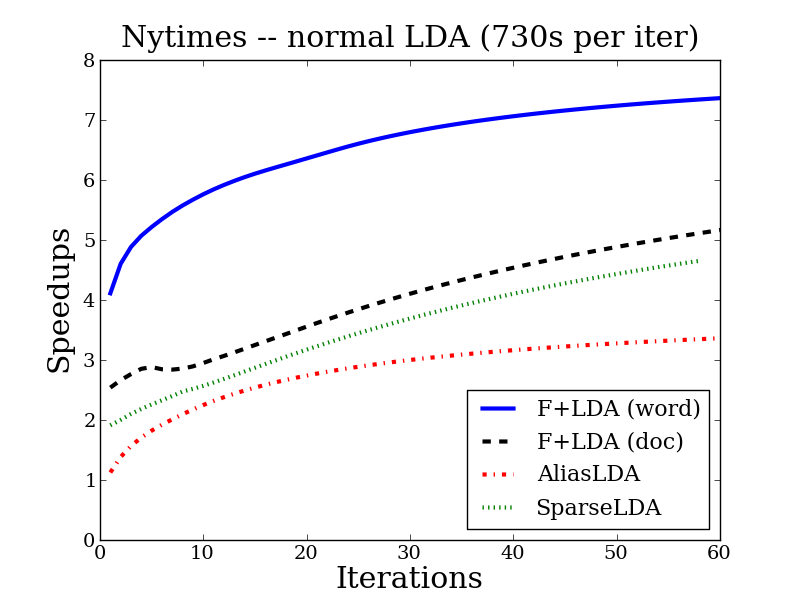}
    \caption{}
    \label{fig:nytimes-speed}
  \end{subfigure}
\end{tabular}
  \caption{(a) and (b) present the convergence speed in terms of number of iterations. 
  (c) and (d) present the sampling speed of each iteration---the y-axis is the speedup over
  the normal LDA implementation which takes $O(T)$ time to generate one sample. 
  We observe all the sampling algorithms have similar convergence speed, while F+LDA(doc) is the fastest
  comparing to other document-wise sampling approaches. Also, F+LDA(word) is faster than
  F+LDA(doc) for larger datasets, which confirms our analysis in Section \ref{sec:treelda}. 
  }
  \label{fig:compare_sampling}
\end{center}
\vspace{-20pt}
\end{figure*}

\subsection{Related Work}
\label{sec:RelatedWork}

Unlike the situation in the serial case, the latest values of
$n_{z,*,w}$ and $n_{z,*,*}$ can be distributed among different machines
in the distributed setting. The existing parallel approaches focus on
development of mechanism to communicate these values.  Next, we briefly
review two approaches for parallelizing CGS in distributed setting:
AdLDA \cite{NewAsuSmyWel09} and Yahoo!\ LDA \cite{SmoNar10}. In both
approaches, each machine has a local copy of the {\em entire}
$n_{z,*,w}$ and $n_{z,*,*}$.  AdLDA uses a bulk synchronization to
update its local copy after each iteration. At each iteration, each
machine just uses the snapshot from last synchronization point to
conduct Gibbs sampling.  On the other hand, Yahoo!\ LDA creates a
central parameter server to maintain the latest values for $n_{z,*,w}$
and $n_{z,*,*}$. Every machine asynchronously communicates with this
machine to send the local update to the server and get new values to
update its local copy. Note that the communication is done
asynchronously in Yahoo!\ LDA to avoid expensive network locking. The
central idea of Yahoo!\ LDA is that modest stale values would not affect
the sampler significantly. Thus, there is no need to spend too much
effort to synchronize these values. Note that for these two approaches,
both values of $n_{z,*,w}$ and $n_{z,*,*}$ used in the Gibbs sampling
could be stale. In contrast, our proposed Nomad LDA has the following
advantages:
\begin{compactitem}
\item No copy of the entire $n_{z,*,w}$ is required in each machine.
\item The value of $n_{z,*,w}$ used in the Gibbs sampling is always
  up-to-date in each machine.
\item The computation is both asynchronous and decentralized.
\end{compactitem}

Our Nomad LDA is close to a parallel approach for matrix
completion~\cite{YunYuHsietal13} in that they also utilized the concept
of nomadic variables. However, the application is completely different.
\cite{YunYuHsietal13} concentrate on parallelizing stochastic gradient
descent for matrix completion. The access graph for this problem is a
bipartite graph, and there is no variable that needs to be synchronized
across processors. Consequently their algorithm is simpler than
Nomad LDA. 
% Each stochastic gradient descent update is associated
% to an edge, and only values in the two end-nodes of this edge are
% changed. The access graph of collapsed Gibbs sampling for LDA, on the
% other hand, is a hyper graph with multi-hyperedges. Further, there is
% one node whose values would be changed in every Gibbs sampling update.
\begin{figure*}[t]
  \begin{subfigure}[t]{0.33\textwidth}
    \includegraphics[width=\linewidth]{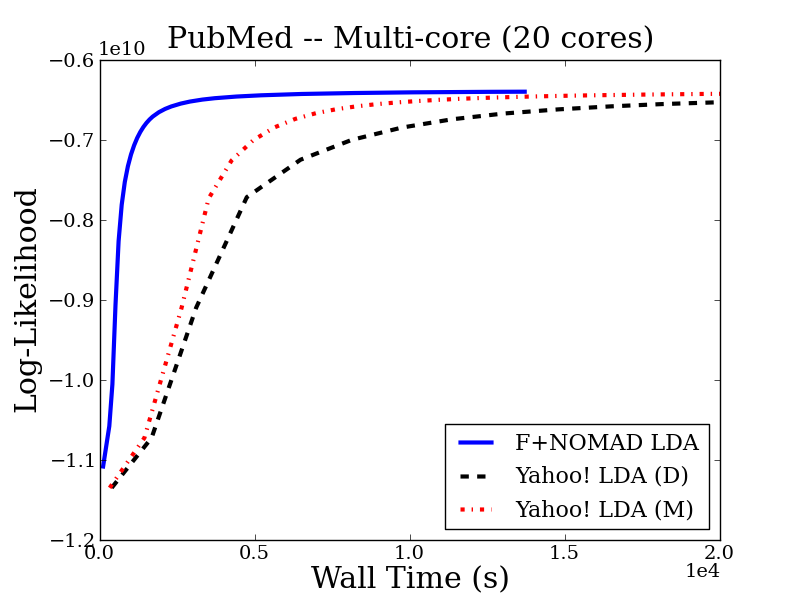}
    \caption{}
    \label{fig:pubmed-multicore}
  \end{subfigure}
  \begin{subfigure}[t]{0.33\textwidth}
    \includegraphics[width=\linewidth]{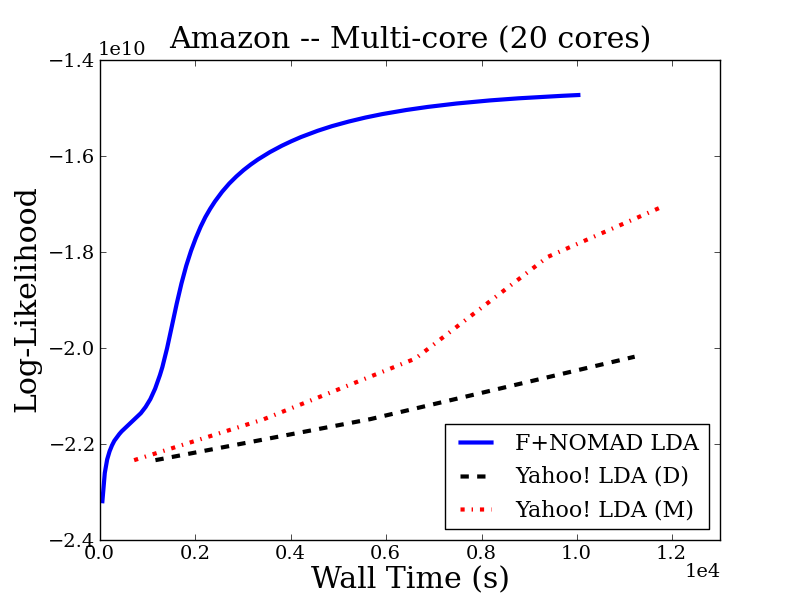}
    \caption{}
    \label{fig:amazon-multicore}
  \end{subfigure}
  \begin{subfigure}[t]{0.33\textwidth}
    \includegraphics[width=\linewidth]{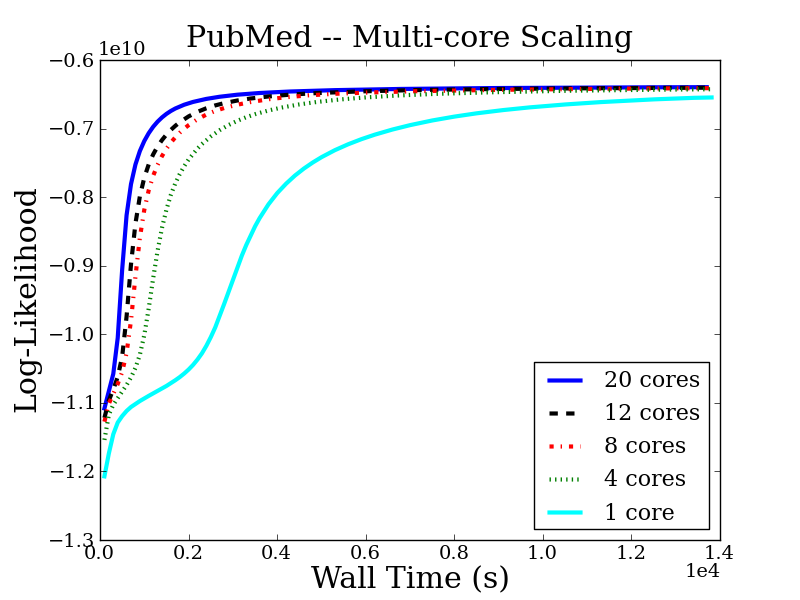}
    \caption{}
    \label{fig:pubmed-scaling}
  \end{subfigure}
  \caption{(a) and (b) show the comparison between Nomad LDA and \yahoolda using 20 cores in a single machine. 
  (c) shows the scaling performance of Nomad LDA as a function of number of cores.}
  \label{fig:multicore}
\end{figure*}

\section{Experimental Evaluation}
\label{sec:ExperEval}

In this section we investigate the performance and scaling of our proposed
algorithms. 
We demonstrate that our proposed F+tree sampling method is very efficient 
in handling large number of topics comparing the other approaches in Section \ref{sec:exp_sampling}. 
When the number of documents is also large, in Section \ref{sec:exp_large} we show
our parallel framework is very efficient in multi-core and distributed systems. 

{\bf Datasets. } We work with five real-world large datasets---Enron, NyTimes, PubMed, Amazon, and UMBC. 
The detailed data set statistics are listed in Table \ref{tab:data}. 
Among them, Enron, NyTimes and PubMed are bag-of-word datasets
in the UCI
repository\footnote{https://archive.ics.uci.edu/ml/datasets/Bag+of+Words}.  
%It contains around 8.2 million documents, and each document on the
%average contains around 90 words which makes it a test corpus with 737
%million words. We chose PubMed because it is used to demonstrate scaling
These three datasets have been used to demonstrate the scaling behavior of 
topic modeling algorithms in many recent papers~\cite{AsuWelSmyTeh09, SmoNar10, LiAhmRavSmo14}. 
In fact, the PubMed dataset stretches the capabilities of many implementations. For instance, we
tried to use LDA code from
\url{http://www.ics.uci.edu/~asuncion/software/fast.htm}, but it could
not handle PubMed.

To demonstrate the scalability of our algorithm, we use two more large-scale 
datasets---Amazon and UMBC. 
The Amazon dataset consists of approximately 35 million product reviews
from Amazon.com, and was downloaded from the Stanford Network Analysis
Project (SNAP) home page. Since reviews are typically short, we split
the text into words, removed stop words, and using Porter
stemming~\cite{Porter80}. After this pre-processing we discarded words
that appear fewer than 5 times or in 5 reviews. Finally, any reviews
that were left with no words after this pre-processing were
discarded. This resulted in a corpus of approximately 30 million
documents and approximately \textbf{1.5 billion} words.

The UMBC WebBase corpus is downloaded from 
\url{http://ebiquity.umbc.edu/blogger/2013/05/01/}. 
It contains a collection of pre-processed paragraphs from the Stanford
WebBase\footnote{Stanford WebBase project: \url{http://dbpubs.stanford.edu:8091/~testbed/doc2/WebBase/}} crawl on February 2007. The original dataset has approximately 40 million 
paragraphs and 3 billion words. We further processed the data by stemming and 
removing stop words following the same procedure in LibShortText~\cite{HFY13a}. 
This resulted in a corpus of approximately \textbf{1.5 billion} words. 

%Our third dataset Usenet is from
%\url{http://www.psych.ualberta.ca/~westburylab/downloads/usenetcorpus.download.html}. It
%contains a minimally pre-processed set of public USENET postings from
%47,860 English language, non-binary-file news groups. This corpus was
%collected between Oct 2005 and Jan 2011. We discarded any posts which
%contained fewer than 50 words. This resulted in a dataset which contains
%15 million documents and approximately \textbf{5.2 billion} words. It is
%roughly 7 times larger than PubMed, and to the best of our knowledge one
%of the largest publicly available datasets. See Table\ref{tab:data} for
%dataset details.

\begin{table}
  \caption{Data statistics.}
  \label{tab:data}
  \vspace{-20pt}
  \begin{center}
    \resizebox{8.5cm}{!}{
    \begin{tabular}{lrrr}
      &  \# documents ($I$) &  \# vocabulary ($J$) &  \# words \\
      \hline
      Enron  & 37,861 & 28,102 & 6,238,796 \\
      NyTimes & 298,000 & 102,660 & 98,793,316 \\
      PubMed & 8,200,000  & 141,043  & 737,869,083 \\
      Amazon & 29,907,995 & 1,682,527 & 1,499,602,431 \\
      UMBC & 40,599,164 & 2,881,476 & 1,483,145,192 \\
 %     Usenet & 15,190,850 &  15,912,790 & 5,240,447,114 \\
    \end{tabular}
    }
  \end{center}
  \vspace{-20pt}
\end{table}

{\bf Hardware. } The experiments are conducted on a large-scale 
parallel platform at the Texas Advanced Computing Center (TACC), Maverick\footnote{\url{https://portal.tacc.utexas.edu/user-guides/maverick}}. 
Each node contains 20 Intel Xeon E5-2680 CPUs and 256 GB memory. Each job can
run on at most 32 nodes (640 cores) for at most four hours.

{\bf Parameter Setting. } 
Throughout the experiments we set the hyper parameters
$\alpha=50/T$ and $\beta=0.01$, where $T$ is number of topics. Previous papers showed that this parameter
setting gives good model qualities \cite{GH08a}, and  
many widely-used software such as Yahoo! LDA and Mallet-LDA also use this
as the default parameter setting.
To test the performance when dealing a large number of topics, we set $T=1024$ in all
the experiments. Our experimental codes are available in 
\centerline{\url{http://www.cs.utexas.edu/~rofuyu/exp-codes/nomad-lda.tgz}.}

{\bf Evaluation. } Our main competitor is \yahoolda in large-scale distributed setting. 
To have a fair comparison with \yahoolda, we use the same training likelihood routine to evaluate the 
quality of model (see eq. (2) in~\cite{SmoNar10} for details). 
\begin{figure*}[thb]
  \begin{center}
  \begin{subfigure}[t]{0.4\textwidth}
    \includegraphics[width=\linewidth]{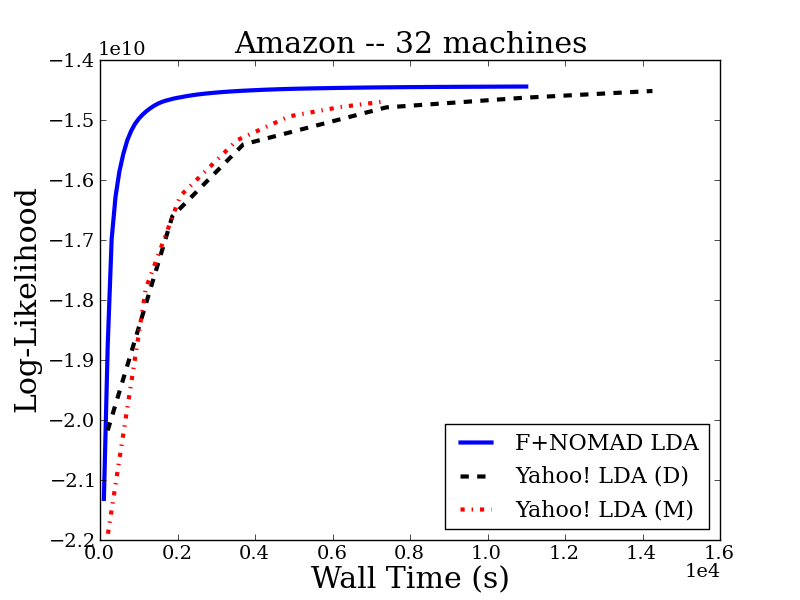}
    \caption{} % Speed comparison}
    \label{fig:amazon-distributed}
  \end{subfigure}
  \begin{subfigure}[t]{0.4\textwidth}
    \includegraphics[width=\linewidth]{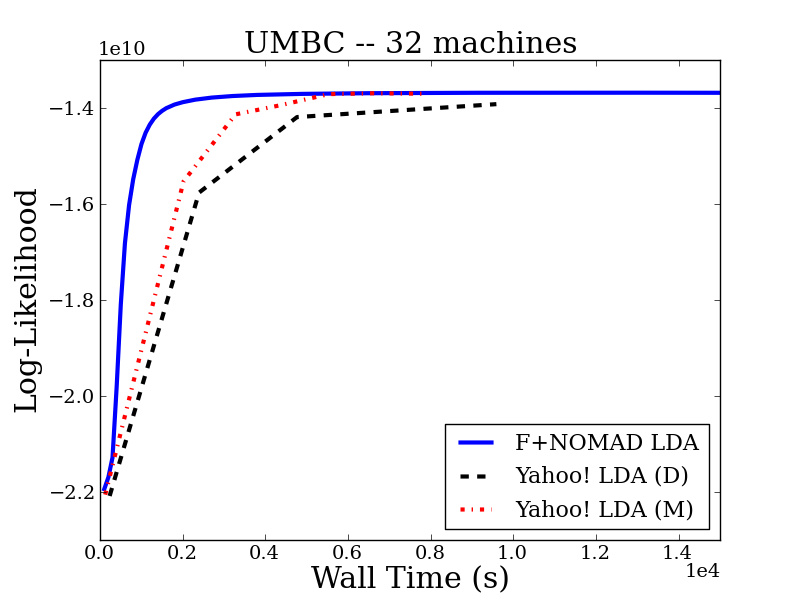}
    \caption{} % Convergence comparison}
    \label{fig:umbc-distributed}
  \end{subfigure}
  \caption{The comparison between F+Nomad LDA and Yahoo!\ LDA on
    32 machines with 20 cores per machine. }
  \label{fig:distributed}
\end{center}
\vspace{-10pt}
\end{figure*}

\subsection{Comparison of sampling methods: handling large number of topics}
\label{sec:exp_sampling}
In this section, we compare various sampling strategies used for
LDA in the serial setting. 
We include the following sampling strategies into the comparison (see Section \ref{sec:ProposedSampling} for details): 
\begin{enumerate}
  \item F+LDA: our proposed sampling approach. We consider both document-wise and word-wise sampling order, denoted by F+LDA(doc) and F+LDA(word) respectively. 
  \item Sparse LDA: the approach that uses linear search on PDF to conduct sampling with document-wise sampling order. This approach is used in Yahoo! LDA and Mallet-LDA. 
  \item Alias LDA: the approach that uses alias method to do the sampling with document-wise sampling order.  
    This approach is proposed very recently in \cite{LiAhmRavSmo14}. 
  \end{enumerate}
To have a fair comparison focusing on different sampling strategies, we implemented the above three approaches under the same data structure. 
We use two smaller datasets---Enron and NyTimes to conduct the experiments. 
Note that \cite{LiAhmRavSmo14} also conducts the comparison of different sampling approaches using these
two datasets after further preprocessing.
Figure \ref{fig:compare_sampling} presents the comparison results.

We first compare the F+LDA(doc), Sparse LDA, and Alias LDA, where all of the three approaches
have the same document-wise sampling ordering. 
F+LDA(doc) and Sparse LDA follow the exact sampling distribution of the normal Gibbs sampling; 
as a result, we can observe in Figure \ref{fig:enron-conv} and \ref{fig:nytimes-conv} that they have the same convergence speed
in turns of number of iterations. 
On the other hand, Alias LDA converges slightly slower than other approaches 
because it does not sample from the exact same distribution. 
In terms of efficiency, Figure \ref{fig:enron-speed} and \ref{fig:nytimes-speed}
indicates that F+LDA(doc) is faster than Sparse-LDA and Alias-LDA, which confirms our
analysis in Section \ref{sec:ProposedSampling}. 

Next we compare the performance of document-wise and word-wise sampling for F+LDA. 
Figure \ref{fig:enron-conv} and \ref{fig:nytimes-conv} indicate that both orderings
give similar convergence speed. 
As discussed in Section \ref{sec:treelda}, using the F+tree sampling approach, the word-wise ordering is expected
to be faster than document-wise ordering as the number of documents increases. This phenomenon 
is confirmed by our experimental results in Figure \ref{fig:enron-speed} and \ref{fig:nytimes-speed}
as F+LDA(word) is faster than F+LDA(doc) on the NyTimes dataset, 
which has a larger number of documents comparing to Enron.
The experimental results also justify our use of word-wise sampling when applying the Nomad approach
in multi-core and distributed systems.

\subsection{Multi-core and Distributed Experiments}
\label{sec:exp_large}
Now we combine our proposed F+tree sampling strategy with the nomadic parallelization
framework. This leads to a new F+Nomad LDA sampler that can handle huge problems
in multi-core and distributed systems. 

%{\bf Implementation Issues. }

\subsubsection{Competing Implementations. } We compare our algorithm against \yahoolda for 
three reasons: a) It is one of the most efficient open
source implementations of CGS for LDA, which scales to large datasets.
b) \cite{SmoNar10} claim that \yahoolda outperforms
other open source implementation such as AD-LDA~\cite{NewAsuSmyWel09}
and PLDA~\cite{WanBaiStaCheetal09}. c) \yahoolda uses a parameter
server, which has become a generic approach for distributing large-scale
learning problems. It is therefore interesting to see if a different
asynchronous approach can outperform the parameter server on this
specific problem. 

\yahoolda is a disk-based implementation that assumes
the latent variables associated with tokens in the documents are
streamed from disk at each iteration. To have a fair comparison, 
in addition to running \yahoolda on normal disk (denoted by \yahooldaD), 
we further ran it on the tmpfs file system~\cite{PS90a} which resides on RAM
for the intermediate storage used by \yahoolda. 
This way we eliminate the cost of disk I/O, and can make a fair
comparison with our own code which does not stream data from disk; we use \yahooldaM to denote
this version.  
%We observe that \yahooldaM takes much less time to conduct one iteration comparing to \yahooldaD, 
%but it usually has slower convergence in terms of log-likelihood, therefore
%
%To have a fair comparison with Yahoo!\ LDA, we use the
%same training likelihood routine to evaluate the quality of the model.
%

\subsubsection{Multi-core Experiments}
\label{sec:MulticoreExperiments}

Both F+Nomad LDA and Yahoo!\ LDA support parallel computation on a single
machine with multiple cores. Here we conduct experiments on 
two datasets, Pubmed and Amazon, and the comparisons are presented 
in Figure \ref{fig:multicore}. 
As can be seen from
Figure~\ref{fig:pubmed-multicore} and \ref{fig:amazon-multicore}, F+Nomad LDA handsomely outperforms 
both memory and disk version of Yahoo!\ LDA, 
and gets to a better quality solution within the same
time budget. Given an desired model quality, F+Nomad LDA is approximately
4 times faster than Yahoo!\ LDA.

Next we turn out attention to the scaling of F+Nomad LDA as a function of
the number of cores. In Figure~\ref{fig:pubmed-scaling} we plot
the convergence of F+Nomad LDA as the number of cores is varied. Clearly,
as the number of cores increases the convergence speed is better. 
%To
%study if the scaling is linear, in
%Figure~\ref{fig:pubmed-multicore-machine}, we plot log-likelihood as a
%function of CPU time which is defined as number of cores $\times$
%elapsed time. If the algorithm scales linearly, then all the curves
%should overlap. There is a clear gap when going from one core to 4
%cores. This is to be expected since a multi-core algorithm has overheads
%and is usually constrained by the memory bandwidth. There is some
%slowdown when going from 4 to higher number of cores, but 
%this does not seem to affect convergence asymptotically. 

\subsubsection{Distributed Memory Experiments}
\label{sec:DistrMemoryExper}

In this section, we compare the performance of F+Nomad LDA and Yahoo!\
LDA on two huge datasets, Amazon and UMBC, in a distributed memory setting.  The number
of machines is set to 32, and the number of cores per machine is 20.  As
can be seen from Figure~\ref{fig:distributed}, F+Nomad LDA dramatically
outperforms both memory and disk version of Yahoo!\ LDA on this task and obtains significantly better
quality solution (in terms of log-likelihood) within the same wall clock time. 
%Next, we investigate
%how the log-likelihood changes as a function of the number of
%iterations. This is important because the sequence of latent variable
%updates performed by Yahoo!\ LDA is different than the he one used by
%Nomad LDA. We want to study if this has an impact on the convergence
%speed. Results can be seen in Figure~\ref{fig:amazon-conv1}.  In terms
%of throughput Nomad LDA can process 325 million words per second, while
%Yahoo!\ LDA is able to process only 41 million words per
%second\footnote{The corresponding numbers for PubMed are 164 million and
%  55 million respectively.}.
%
%Next we study the scaling behavior of Nomad LDA. In
%Figure~\ref{fig:amazon-real}, we observe that Nomad LDA enjoys almost
%linear scaling as we move from 80 cores (4 machines, 20 cores per
%machine) to 640 cores (32 machines, 20 cores per machine).

\section{Conclusions}

In this paper, we present a novel F+Nomad LDA algorithm that can handle
large number of topics as well as large number of documents. In order to
handle large number of topics we use an appropriately modified Fenwick
tree. This data structure allows us to sample from and update a
$T$-dimensional multinomial distribution in $O(\log T)$ time. In order
to handle large number of documents, we propose a novel asynchronous and
non-locking parallel framework, which leads to a good speedup in
multi-core and distributed systems.  The resulting algorithm is faster
than Yahoo\! LDA and is able to handle datasets with millions of
documents and billions of words. In future work we would like to include
the ability to stream documents from disk, just like Yahoo\! LDA. It is
also interesting to study how our ideas can be transferred to other
sampling schemes such as CVB0. 

\clearpage
%\section{Acknowledgments}
\bibliographystyle{abbrv}
\bibliography{nomad-lda}
\end{document}